%% file: manuscript.tex
\documentclass[fleqn,usenatbib]{mnras}

\include{newcommands}

\usepackage{newtxtext,newtxmath,lineno}

\usepackage[T1]{fontenc}
\usepackage{ae,aecompl}

\usepackage{graphicx}	
\usepackage{amsmath}	
\usepackage{amssymb}	
\usepackage{tikz,xcolor,hyperref}
\usepackage{lineno}
\usepackage{multirow}
\usepackage{booktabs}

\include{newcommand_gen}
\newcommand{\thestar}{2M0632}

\title[An M Dwarf ``Dipper" Star with a Long-Lived Disk]{Planetesimals Around Stars with \tess\ (PAST): II. An M Dwarf ``Dipper" Star with a Long-Lived Disk in the \emph{TESS} Continuous Viewing Zone}

\author[Gaidos et al.]{
Gaidos, Eric,$^{1,2,3}$
\thanks{E-mail: gaidos@hawaii.edu} 
Andrew W. Mann,$^{4}$
B\'{a}rbara Rojas-Ayala,$^{5}$
Gregory A. Feiden,$^{6}$
\newauthor
Mackenna L. Wood,$^{4}$
Suchitra Narayanan,$^{7}$
Megan Ansdell,$^{8,9}$
Tom Jacobs,$^{9}$
\newauthor
Daryll LaCourse$^{9}$
\\
$^{1}$Department of Earth Sciences, University of Hawai`i at M\={a}noa, Honolulu, HI  96822, USA\\
$^{2}$Center for Space and Habitability, University of Bern, Gesellschaftsstrasse 6, 3012 Bern, Switzerland\\
$^{3}$Institute for Astrophysics, University of Vienna, T\"{u}rkenschanzstrasse 17, 1180 Vienna, Austria\\
$^{4}$Department of Physics and Astronomy, University of North Carolina at Chapel Hill, Chapel Hill, NC 27599-3255, USA\\
$^{5}$Instituto de Alta Investigaci\'on, Universidad de Tarapac\'a, Casilla 7D, Arica, Chile\\
$^{6}$Department of Physics \& Astronomy, University of North Georgia, Dahlonega, GA 30597, USA\\
$^{7}$Institute for Astronomy, University of Hawai`i at M\={a}noa, Honolulu, HI 96822, USA\\
$^{8}$NASA Headquarters, 300 E Street, N.W., Washington, DC 20546, USA\\
$^{9}$Visual Survey Group\\
}

\date{Accepted 2022 April 25. Received 2022 April 10; in original form 2022 February 4}

\pubyear{2022}

\hypersetup{draft}
\begin{document}
\label{firstpage}
\pagerange{\pageref{firstpage}--\pageref{lastpage}}
\maketitle

\begin{abstract}
Studies of T Tauri disks inform planet formation theory; observations of variability due to occultation by circumstellar dust are a useful probe of unresolved, planet-forming inner disks, especially around faint M dwarf stars.  We report observations of \thestar, an M dwarf member of the Carina young moving group that was observed  by \tess\ over two one-year intervals.  The combined light curve contains $>$300 dimming events, each lasting a few hours, and as deep as 40\% (0.55 magnitudes).  These stochastic events are correlated with a distinct, stable 1.86-day periodic signal that could be stellar rotation.  Concurrent ground-based, multi-band photometry show reddening consistent with ISM-like dust.  The star's excess emission in the infrared and emission lines in optical and infrared spectra, reveal a T Tauri-like accretion disk around the star.  We confirm membership of \thestar\ in the Carina group by a Bayesian analysis of its Galactic space motion and position.  We combine stellar evolution models with \gaia\ photometry and constraints on \teff, luminosity, and the absence of detectable lithium in the photosphere to constrain the age of the group and \thestar\ to 40-60 Myr, consistent with earlier estimates.  \thestar\ joins a handful of long-lived disks which challenge the canon that disk lifetimes are $\lesssim10$\,Myr.  All known examples surround M dwarfs, suggesting that lower X-ray/UV irradiation and slower photoevaporation by these stars can dramatically affect disk evolution.  The multi-planet systems spawned by long-lived disks probably experienced significant orbital damping and migration into close-in, resonant orbits, and perhaps represented by the TRAPPIST-1 system.
\end{abstract}




\section{Introduction}
\label{sec:intro}

Space-based transit surveys, particularly \kepler, have revealed that Earth- to Neptune-size planets are common around other stars \citep{Silburt2015,Hsu2019,Bryson2021}. They also indicate that the occurrence of small planets on close-in orbits increases with decreasing stellar mass \citep{Mulders2015,Hardegree-Ullman2019}.  Due to the low luminosity of these stars, some of these planets could have equilibrium temperatures that are consistent with the stability field of liquid water, and could in principle be habitable if they are rocky and Earth-like.  The planetary systems of nearby M dwarfs represent the most promising targets for characterization of habitable planets and searches for atmospheric biosignatures, e.g. with \jwst\ \citep{Greene2019,Lustig-Yaeger2019a,Tremblay2020}.

M dwarf stars differ significantly from their solar-mass counterparts: do these differences manifest themselves in the properties of their planetary systems?   For example, impacts onto planets in the compact habitable zones of M dwarfs will be at higher velocities and be more erosive of any atmospheres \citep{Lissauer2007}.  These planets will also experience elevated high-energy (X-ray and UV) irradiance \citep{Ansdell2015}, which is expected to heat, inflate, and evaporate atmospheres \citep{Owen2019}.  The slower pre-main sequence evolution of M dwarfs means that planets orbiting in the habitable zone experience prolonged high total irradiation at early times that could drive a runaway greenhouse and loss of H$_2$O by escape of H to space \citep{Luger2015}.  Other differences could arise during planet formation due to differences in the structures \citep{Gaidos2017b}, temperature distribution \citep{Kennedy2007,Mulders2021}, and lifetime \citep{Zawadzki2021} of disks.  There may be other, less obvious but equally profound differences such as the abundance of short-lived radionuclides (e.g., $^{26}$Al) that heat the interiors of planetesimals \citep{Gaches2020}.

Studies of disks around young stellar objects (YSOs) as  counterparts to those which formed the planetary systems detected around older stars offer insight into potential variation along the main sequence.  But the regions of most disks corresponding to the orbits of most known planets ($<1$\,au) cannot be resolved at the distance of the nearest star-forming regions, even by the ALMA mm-wave interferometer.  Near-infrared Interferometers such as GRAVITY can resolve the inner regions of the brightest T Tauri disks, but sample the $u$-$v$ spatial frequency plane less and allow fitting of relatively simple disk models \citep{Perraut2021,Bohn2022}.   Disks around M dwarf stars are even more challenging to study because they are less luminous and host less massive disks \citep{Pascucci2016}.  As a result, comparatively little is known about disks around the lowest-mass stars.  Multiple studies found a trend of increasing disk fraction at a given age (and by infererence disk lifetime) with decreasing stellar mass \citep[e.g.,][]{Luhman2012,Ribas2015}, although this trend may be at least in part the product of detection bias and inappropriate pre-main sequence stellar models \citep{Richert2018}.  Disk mass and accretion rates are observed to approximately scale with stellar mass and its square, espectively \citep[][ and references therein]{Manara2022}, naively suggesting disk lifetime inversely decreases with stellar mass.  \citet{Kastner2016} suggest substantially longer disk lifetime among  M dwarfs related to their intrinsic high-energy luminosity and pre-main sequence lifetime.  In addition, a handful of other long-lived ($>10$ Myr) disks have been identified, all around M dwarfs \citep{Silverberg2020}.  This may be related to low predicted evaporation rates for disks around such stars \citep{Wilhelm2022}.      

One approach to study inner protoplanetary disks is to observe the variability produced by circumstellar material as it occults the star \citep{Herbst1994}.  As is the case for exoplanets, observations of occultations are far more sensitive to circumstellar material than scattering or emission, and can be performed on fainter stars such as M dwarfs.  If a disk is highly inclined towards our line of sight, occultations can be produced by stellar magnetosphere-funneled accretion flows from the inner disk edge onto the star \citep{Bouvier2003,Bodman2017}, and vertical structures that are the result of instabilities such as the Rossby wave instability \citep{Stauffer2015,Ansdell2016a}.  Other mechanisms need not require an inclined disk, or a disk at all, e.g., dust lofted in disk winds \citep{Varga2017,Fernandes2018}, gravitationally bound clumps of planetesimals \citep{Ansdell2016a},  disintegrating comet-like planetesimals \citep{Kennedy2017,Ansdell2019a}, or evaporating planets \citep{Rappaport2012,Sanchis-Ojeda2015}.  Such occultations are opportunities to probe the properties and composition of dust and gas close to the star through wavelength-dependent scattering and polarization \citep{Natta1997,Bouvier2014}, and atomic or molecular line absorption \citep{Sorelli1996,Zhang2015,Gaidos2019a}.     

Although dimming due to occultation by circumstellar dust has been studied from the ground for decades, e.g. among UX Orionis-type variables and a few low-mass stars \citep[e.g.,][]{Herbst1994,Bouvier1999}, recent space-based photometry of ($\lesssim$10~Myr) star-forming regions by \emph{CoRoT}, \emph{Spitzer} and \ktwo\ has greatly expanded the number and diversity of stars known to exhibit dimming.  Notably, many ``dipper" stars exhibit quasi-periodic or stochastic dimming that is on the timescale of one day  \citep{Alencar2010,Morales2011,Cody2014,Ansdell2016a}; this photometric behavior is difficult to detect and monitor from the ground.  The \emph{Transiting Exoplanet Survey Satellite} (\tess) is obtaining precise photometry over nearly the entire sky with a cadence of $\leq$30\,min for intervals of at least 27 days \citep{Ricker2014}.  This permits monitoring of and discovery of dimming events in stellar clusters and dispersed young moving groups (YMGs) with an even great range of ages \citep{Gaidos2019b}.  The \gaia\ astrometric mission  \citep{Gaia2016,Lindegren2018} is allowing new members of such YMGs to be identified, as well as new groups themselves \citep{Faherty2018,Tang2019}, and to estimate the ages of these groups by comparison of color-magnitude diagrams to stellar models.   This \gaia-\tess\ synergy promises to revolutionize our understanding of stellar variability caused by orbiting planetesimals and dust.

Here we describe 2M0632, a nearby ($\approx$93 pc) M-dwarf member of the Carina young moving group that exhibits both dipper-like variability as well as all the hallmarks of a T Tauri-like accretion disk.  Its location in the \tess\ Southern Continuous Viewing Zone (SCVZ) means that \tess\ observations have afforded two nearly continuous year-long light curves, providing a high-precision picture of its variability on time scales of minutes to a $\sim$1 year.  The Carina young moving group was previously assigned an age of $\approx$45~Myr \citep{Bell2015}, thus this star along with several previously described systems \citep{Murphy2018,Silverberg2020}, seem to have maintained disks for an order of magnitude longer than the canonical disk lifetime of $\sim$5 Myr \citep{Bell2013}.  In this work, we combine the unparalleled \tess\ dataset of this star with ground-based observations to examine the dimming phenomenon in more detail, investigate the properties of its disk, and exploit the precision of \gaia\ astrometry and photometry to re-assess the age of the host stellar group and this unusual object.   

\section{Observations and Data Reduction}
\label{sec:observations}

\subsection{\tess}

 \tess\ photometer data were retrieved from the Mikulski Archive for Space Telescopes (MAST).  \tess\ light curves are manually inspected by the Visual Survey Group, a team of citizen and academic scientists, for notable phenomena, including ``dipper" stars.  The search is conducted using {\tt LcTools} \citep{Schmitt2019}, a free and publicly available software program that provides a set of applications for efficiently building and visually inspecting large numbers of light curves \citep{Kipping2015}. For more details on the {\tt LcTools} package and the visual survey methodology, see \citet{Rappaport2018}.  \tess\ Input Catalog (TIC) source 167303776 was flagged for its variability by the Visual Survey Group on 3 March 2019, and subsequently identify by \citet{Tajiri2020} as a candidate ``dipper" star using a convolutional neural network trained on a set of eclipsing binary light curves.  TIC~167303776 is associated with the 2MASS infrared source J06320799-6810419 and hereafter is referred to as \thestar. 

\begin{figure*}
	\includegraphics[width=\textwidth]{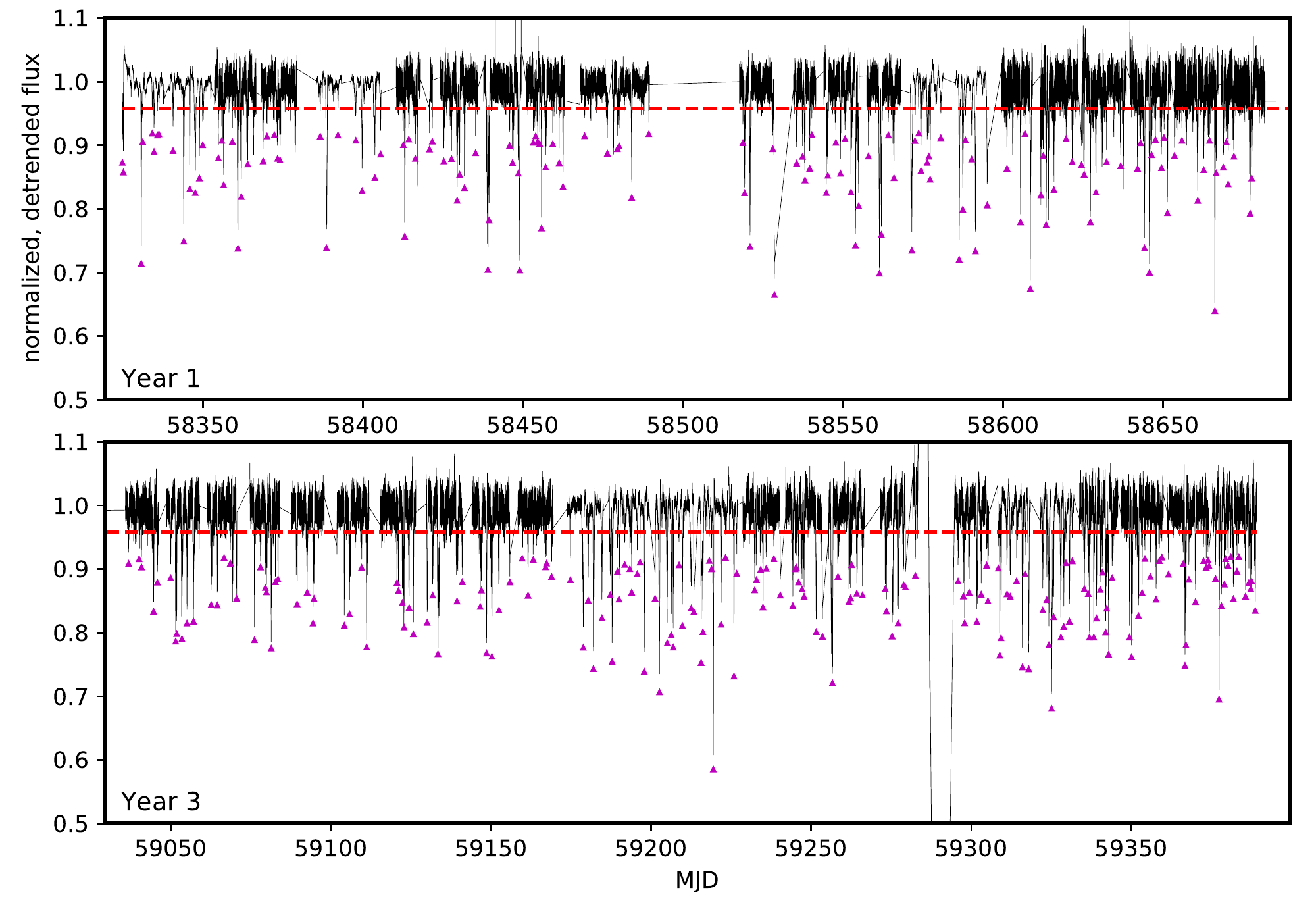}
    \caption{Detrended, normalized composite light curve of \thestar\ assembled from 25 sectors in mission Years 1 and 3.  The large gap in Year 1 is during Sector 7 when the star was not observed.  Smaller gaps at every spacecraft orbit ($\approx$13 days) occur when the spacecraft was oriented for data downlink or re-pointing.   The horizontal dashed line indicates the threshold for dimming events or ``dips", and the magenta points designate systematically identified dips (see Sec. \ref{sec:tess_analysis}).  The anomalies around MJD = 59288 are incompletely-corrected artifacts at the end and beginning of observation Sectors.}
    \label{fig:tess_lc}
\end{figure*}

\thestar\ was observed by \tess\ over the first (Sectors 1-13) and third (Sectors 27-39) years of its mission, with the exception of a 27-day interval (Sector 7).  The star was not selected as a 2-min cadence target in Sectors 1, 3, 10, 32, and 37, thus light curves for those sectors were obtained from the {\tt TESS-SPOC} pipeline analysis of the 30- or 15-min Full Frame Image data \citep{Caldwell2020}. The Pre-search Data Conditional Simple Aperture (PDCSAP) photometry with an optimized 3- or 4-pixel aperture was used.  Slow variation that is an artifact of spacecraft pointing drift and the rotation of the field between sectors was removed by iterative third-order, 300-point Savitzky-Golay filtering \citep{Savitsky1964} of a high- and low-flux-clipped set of points, where the clip thresholds of top 1\% and bottom 40\% were chosen to exclude dimming events based on inspection.  The light curve was normalized by the filtered, clipped curve, the ordered pixel values and clip thresholds recomputed, and the fit repeated iteratively until convergence.    To make the noise properties of the composite light curve, which contains observations at different cadence, more uniform, the detrended version was convolved with a Gaussian with a FWHM of 15 min.  This preserves the structure in the long-cadence data (30 min) while increasing the SNR of the short-cadence (2 min) data.  Figure \ref{fig:tess_lc} shows the resulting normalized, composite light curve.

\subsection{LCOGT photometry}

Observations of \thestar\ were obtained with the 0.4-m telescope network of the Las Cumbres Observatory Global Telescope \citep[LCOGT,][]{Brown2013}. Each telescope is equipped with a 3K $\times$ 2K SBIG 6303 camera with 0\arcsec.571 pixels and a field of view of 29\arcmin.2 $\times$ 19\arcmin.5.  Photometry was obtained through Sloan $gri$ and Pan-STARRS $Z$ filters.  All five 0.4-m telescopes at the austral sites of LCOGT (Sutherland/SAAO, CTIO, and Siding Spring) were used.   The first set of observations between UT 17 October and 23 November 2019 comprised 75 sets of 3 exposures through each of the $griZ$ filters.  Individual exposure times were 300, 300, 270, and 300 sec, respectively.  The second set, obtained between December 3, 2021 and January 21, 2022, consisted of 142 sets of $griZ$ images, each with 600 sec integrations. 

LCOGT images are processed by the {\tt BANZAI} pipeline \citep{McCully2018}, which performs bias removal, flat-fielding, and source identification and aperture photometry using the {\tt SExtractor} routines \citep{Bertin1996}.  Along with photometry in several apertures of fixed diameters, fluxes are also measured within an ``optimal" elliptical Kron aperture \citep{Kron1980}.  Although Kron photometry was originally developed for galaxy photometry, stars in LCOGT images are often non-circular due to imperfect tracking, focus, or astigmatism, and we adopted that approach for our analysis.

Relative, time-series photometry was calculated using the procedures more fully described in Narayanan et al., in prep.; we highlight the major features of the algorithm here. For the sources in each band-pass, including \thestar, photometry was calculated by matrix solution of a set of linear equations describing the relation between instrumental magnitudes $m_{ij}^{\rm inst}$ of the $j$th observation of the $i$th star with instrument $k$ (a unique site/enclosure/telescope/camera index), apparent magnitudes $m_i^{\rm app}$, observation zero-points $Z_j$, a second-order extinction coefficient $\beta$, and an instrument-specific color-term $\gamma$,
\begin{equation}
\label{eqn:photometry}
    m_{ij}^{\rm inst} = m_i^{\rm app} + Z_j + \beta C_i (X_i - 1) + \gamma_{k(j)} C_i ,  
\end{equation}
Note that since this is \emph{relative}, not absolute, photometry, terms that are invariant with observation for any given star, such as a color term, are ignored.  We adopted the \gaia\ $B_P-R_P$ color for $C$.   When solving for the coefficients of Eqn. \ref{eqn:photometry}, we only use those sources with peak pixel value $<20000$ (to avoid saturation), and belong to stars with \gaia\  fractional variability $\sigma<$0.3\%, where:
\begin{equation}
    \label{eqn:variability}
    \sigma \approx \frac{\sigma_f}{f}\sqrt{\frac{N}{8.86}}.
\end{equation}
Here $f$ ({\tt photo\_g\_mean\_flux}) and $\sigma_f$ ({\tt phot\_g\_mean\_flux\_error}) are the \gaia\ source brightness (counts sec$^{-1}$) and its standard error, and $N$ is the number of individual photometric observations ({\tt photo\_g\_n\_obs}).  \gaia\ makes multiple photometric measurements during the brief passage or transit of a star across the detector field of view; the star will not vary significantly during a transit, thus the effective number of epochs at which \gaia\ measures variability is reduced by the mean number of observations per transit, which is 8.86 \citep{Jordi2010}.  The variance of each individual star is calculated after solution of Eqn. \ref{eqn:photometry}, and a small but arbitrary (here 3\%) fraction of the most variable stars are removed from the sample before re-calculating the photometry.  The median error in the zero-points is used as a metric for the quality of  the solution.  As more of the most discrepant stars are removed, the error in zero-points decreases, but beyond some optimal point when the median zero-point error is minimized, the decrease in sample size results in increasing error.  We adopted the solution at the minimum for calculating light curves for \thestar.

In a subset of the images obtained with one particular detector, the image of \thestar\ fell on a ``hot" pixel which produced errors in the photometry.  These data were excluded based on proximity of the PSF centroid to this pixel.  Many $g$-band images -- usually the first in a series of exposures -- suffered from residual telescope shake and poor tracking after a telescope slewed to a target.  This smeared stellar images and made detection and accurate photometry of \thestar, already faint in this band-pass, problematic.  To mitigate this, we excluded data from images for which the seeing (L1FWHM, which is the median PSF FWHM for all sources in the image) is $>$3\arcsec.  (Here, ``seeing'' includes the effects of focus and tracking problems.)      

\begin{figure*}
	\includegraphics[width=\textwidth]{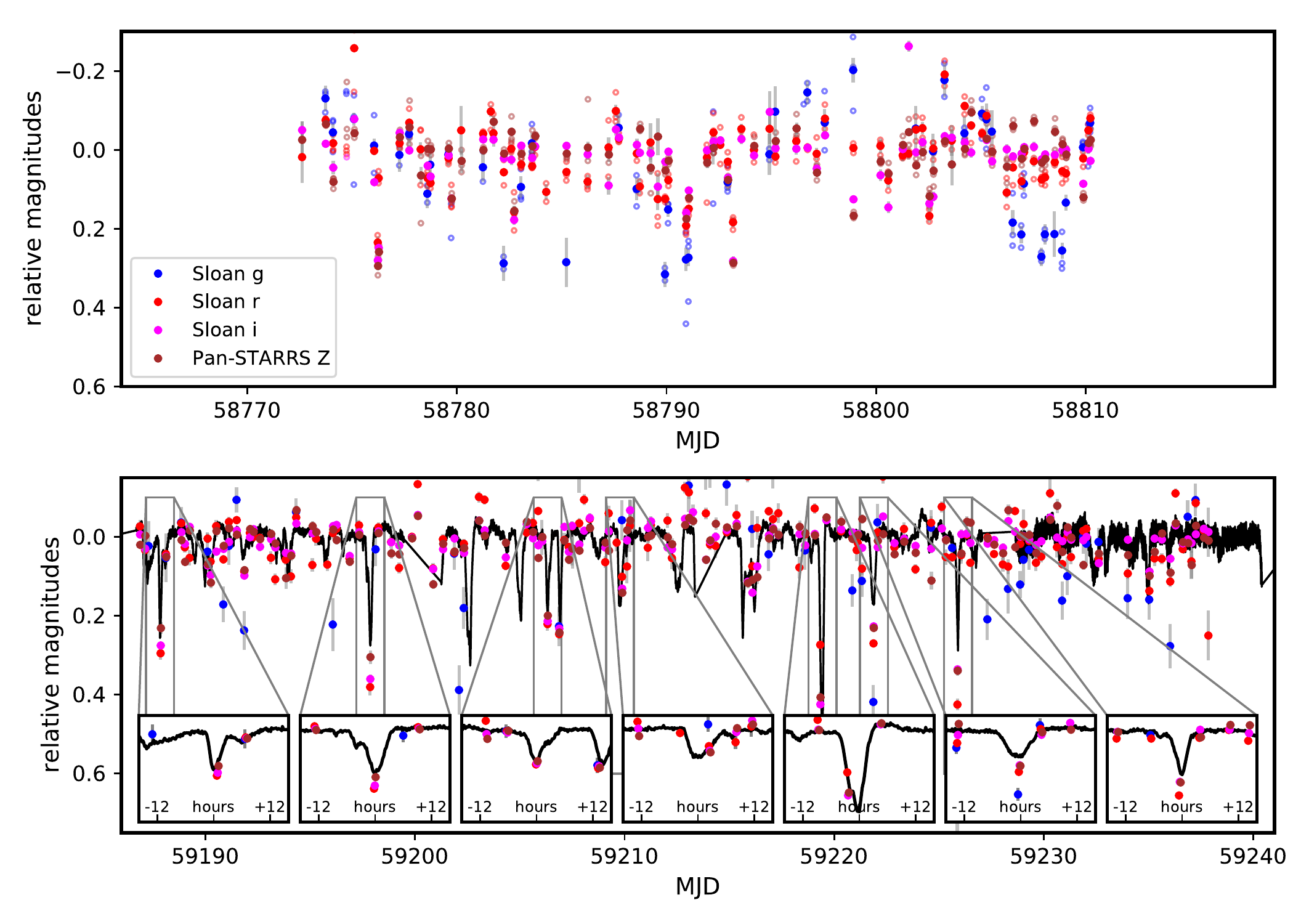}
    \caption{LCOGT multi-bandpass photometry of \thestar\  obtained during two campaigns.  The second data set was obtained contemporaneously with \tess\ observations, which are also plotted.  Insets show 32-hour intervals centered on some of the deepest dips observed by both \tess\ and LCOGT.  Open points represent individual measurements in a series obtained at a single epoch; in these cases the filled points represent the mean.}
    \label{fig:lco}
\end{figure*}

Figure \ref{fig:lco} plots the LCO $griz$ photometry along with the \tess\ light curve.  To remove as far as possible uninteresting offsets between the datasets, the median value after exclusion of the faintest 16\% of points (those likely to be obtained when the star in the dimm state) was subtracted.  The threshold of 16\% was based on an analysis of the \tess\ light curve (Sec. \ref{sec:tess_analysis}).  The two LCO campaigns are plotted in separate panels in Fig. \ref{fig:lco}; the second contains the concurrent \tess\ observations with insets showing the details of some dips observed with both \tess\ and LCO. 

\subsection{Spectroscopy}

On the night of 2019 December 2 (UT), we observed \thestar\ with the Goodman High-Throughput Spectrograph \citep{Goodman_spectrograph} on the Southern Astrophysical Research (SOAR) 4.1\,m telescope atop Cerro Pach\'{o}n, Chile. We took all exposures using the red camera, the 1200 lines mm$^{-1}$ grating, the M5 setup, and the 0\arcsec.46 slit rotated to the parallactic angle. This setup provides a nominal resolution of $\lambda/\Delta \lambda \approx$5000 spanning 6350--7500\AA, although we found the effective resolution to be closer to 3000 due to drift of the wavelength solution (see below).  We took five exposures, each with 600\,second integration times. We took a standard set of calibration data during the daytime (flats and biases) as well as a set of Ne arcs immediately before observing the target. 

\begin{figure*}
	\includegraphics[width=\textwidth]{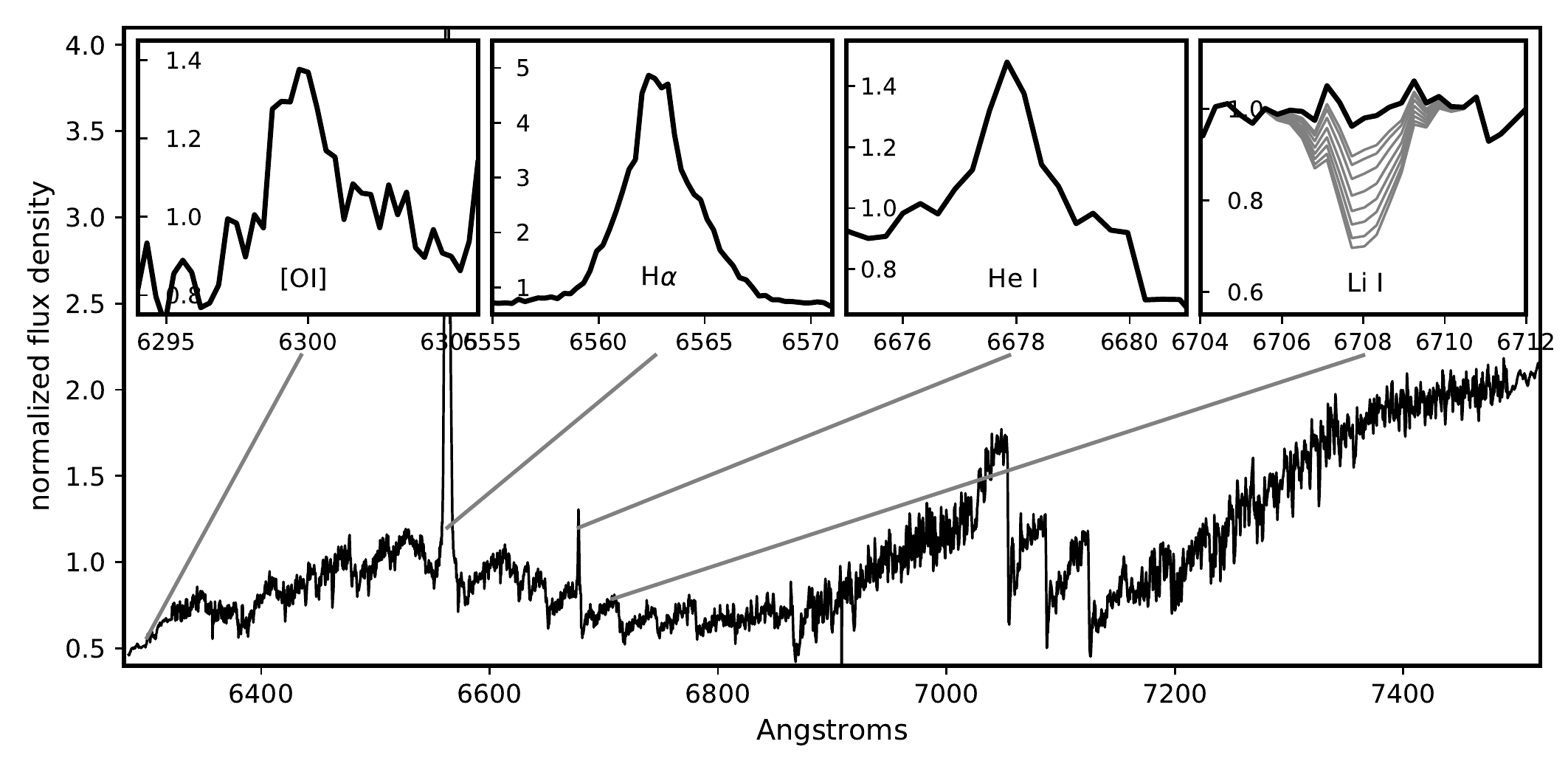}
    \caption{Normalized SOAR/Goodman echelle spectrum of \thestar, with insets showing emission lines in [OI] at 6300\AA, H$\alpha$ at 6563\AA, triplet He\,I at 6678\AA, and the resonant double of Li\,I at 6708\AA\ (spectra are locally normalized for the insets).  The combination of five individual spectra is plotted except for the [OI] region, for which a single spectrum is plotted.  To compare with the non-detection of Li\,I, the grey curves are simulated spectra with lines for $A(Li)$ of 0, 0.5, 1, 1.5, 2, 2.5, 3., and 3.3 (A(H) = 12), using the curve of growth of \citet{Palla2007} and assuming a Gaussian line profile with the width set by the ideal instrument resolution.}
    \label{fig:soar}
\end{figure*}

Using custom scripts, we performed bias subtraction, flat fielding, optimal extraction of the target spectrum, and mapping pixels to wavelengths using a fifth-order polynomial derived from the Ne lamp spectra. We then stacked the five extracted spectra using the robust weighted mean. The stacked spectrum had a signal-to-noise ratio (SNR) $>$100 over most of the wavelength range.  We measured the radial velocity of \thestar\ by cross-correlating the Goodman spectrum against a series of radial-velocity template spectra from \citet{Nidever2002}. Due to an issue with the flexure compensation system on SOAR, the wavelength solution can shift during exposures. This caused some loss in spectral resolution and could impact the inferred radial velocities by as much as 30 km\,s$^{-1}$ (based on tests with radial velocity standards). To correct for this, we derived a linear fit to wavelengths of the OH sky emission lines and O$_2$ absorption lines in each extracted spectra (prior to background subtraction). The shifts are larger near the edges of the spectrum, but there are too few skylines for a higher-order fix, so we masked out 100\AA\ on each side. The final velocity from our analysis was 21.5 $\pm$ 2.3 km\,s$^{-1}$. The error is based on the cross-correlation and the uncertainty in the skyline fit; however, tests on RV standards suggest the uncertainty could be as high as $\simeq$3~km\,s$^{-1}$.  The [OI] line at 6300\AA\ is at the edge of the first two spectra but falls partially or completely off the edge in the others. For that region, we used only the first spectrum.  The extracted, normalized spectrum is plotted in Fig. \ref{fig:soar}.

\begin{figure*}
	\includegraphics[width=\textwidth]{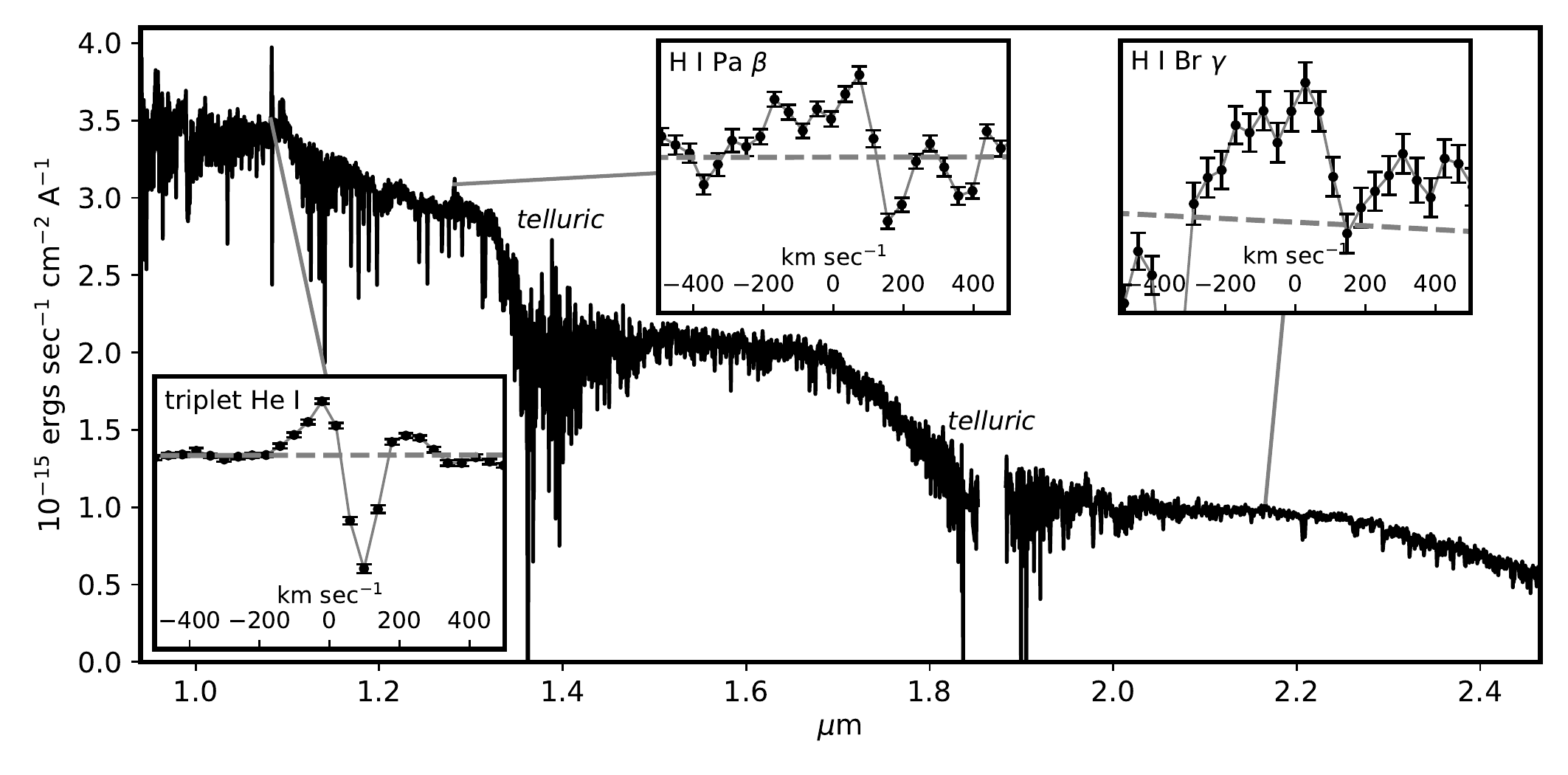}
    \caption{$JHK$ spectrum of \thestar\ obtained with Triple-Spec-4 on the CTIO 4-m telescope, with insets showing the 1.083 \micron\ line of triplet He I and the Paschen $\beta$ and Brackett $\gamma$ lines of H I.  The dashed grey lines are pseudo-continuum levels derived by spline fits to neighboring points.}
    \label{fig:tspec}
\end{figure*}

We obtained infrared spectra of \thestar\ with the TripleSpec-4.1 instrument during its Science Verification stage on the SOAR 4.1-m telescope in April 2019. TripleSpec-4.1 is a cross-dispersed spectrograph featuring a fixed slit of 1\arcsec.1 by 28\arcsec, resulting in spectra of resolution of $\approx$ 3500 and covering a simultaneous wavelength range of 0.8 to 2.47 \micron\ \citep{Schlawin2014}. Using an A-B nod pattern, we obtained four consecutive spectra of \thestar, shifting the star along the slit to perform sky subtraction. We did the same for the A0V star HD~61834, used to correct the \thestar spectra of telluric absorption.  We reduced the data using the version of {\tt SpexTool} for the TripleSpec 4.1 at SOAR\footnotemark\footnotetext{https://noirlab.edu/science/observing-noirlab/observing-ctio/observing-soar/data-reduction/triplespec-data} \citep{Cushing2004}. We constructed a telluric spectrum from the spectrum of HD\,61834 and used this to correct the spectrum of \thestar\ with the routine {\tt xtellcor} \citep{Vacca2003}.  This corrected spectrum is plotted in Fig. \ref{fig:tspec}; insets show details of emission in the lines of triplet He~I at 1.083 \micron\ and H I Pachen-$\beta$ and Brackett $\gamma$.

\subsection{Adaptive optics imaging}
\label{sec:ao}

Adaptive optics (AO) imaging was obtained with the Nasmyth Adaptive Optics Camera Near-Infrared Imager and Spectrograph \citep[NACO;][]{Lenzen2003,Rousset2003} at the UT1 8-m telescope at Paranal on UT 19 January 2019, as part of a survey (program 1101.C-0092(D), PI: R. Launhardt), and we retrieved these data from the ESO Science Archive.  NACO delivered a 14\arcsec\ $\times$ 14\arcsec\ field of view with 13.22 mas pixels.  The single usable image is a 10-sec exposure through the $K_s$ filter at an airmass of 1.4.  A single saturated source appears in the image; the WCS solution locates it 1\arcsec.65 east and 7\arcsec.67 north of the \gaia\ location, but since a 2MASS $K_s$ image shows no comparably bright sources within $\sim$1', we conclude this offset is the boresight error.  There are no detected sources within the image, within 0\arcsec.53 for any position angle or within 0".75 for 75\% of possible position angles.

\section{Analysis}
\label{sec:analysis}

\subsection{Validation and Characterization of transient dimming}
\label{sec:tess_analysis}

\begin{figure}
	\includegraphics[width=\columnwidth]{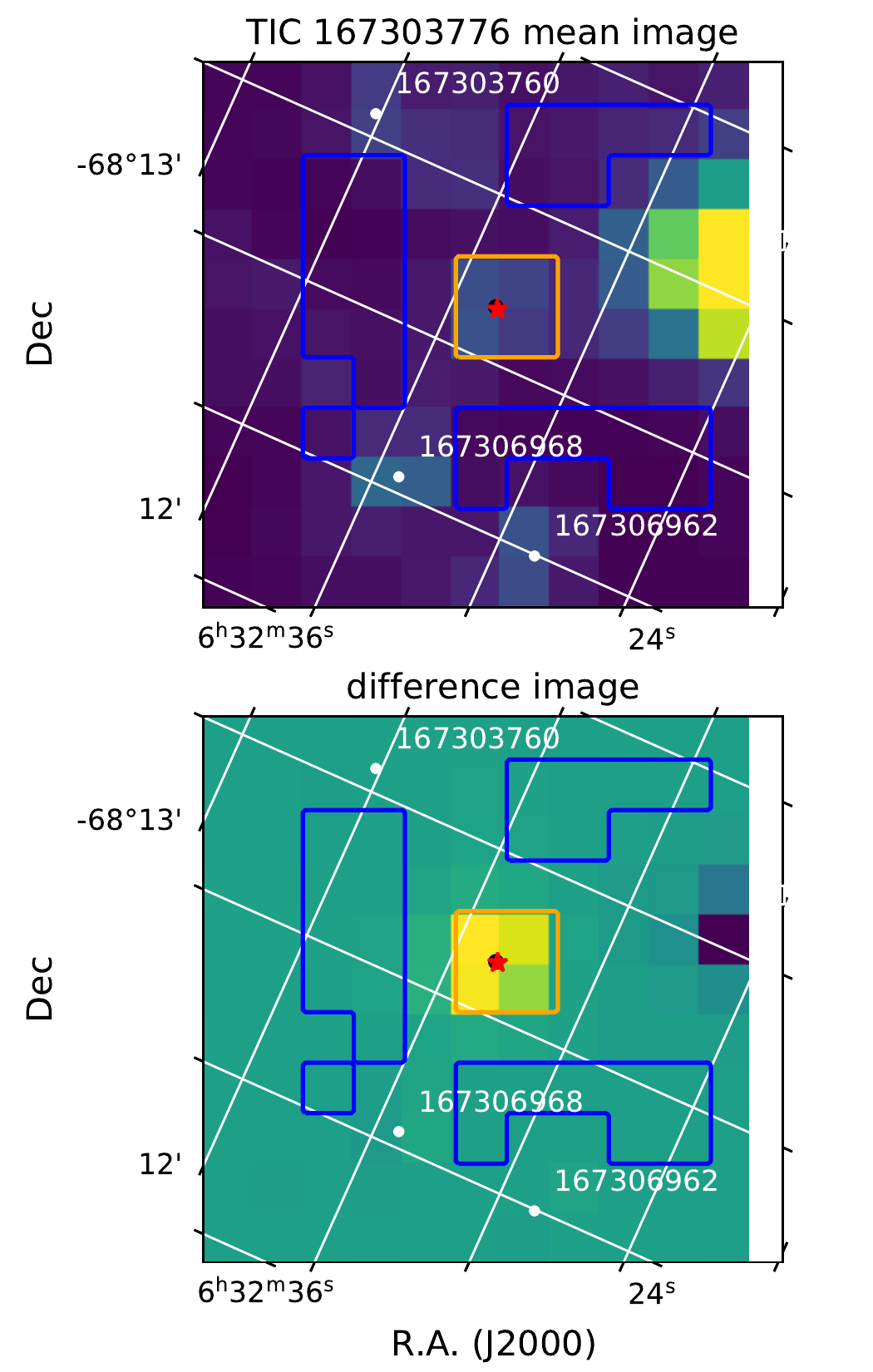}
    \caption{Top: Mean \tess\ photometer image from Sector 2 observations of \thestar.  The black point marks \thestar; other stars brighter than \tess\ magnitude $T = 15$ are marked by white points.  The red star is the flux centroid within the aperture.  Source and background apertures used to construct the light curve in this Sector are outlined by orange and blue lines, respectively.  Bottom: Difference image constructed by taking the mean flux inside of dimming events (defined to occur when the flux drops $>$4.15\% relative to the median) minus the mean flux outside.  This shows that the dimming signal coincides with the location of the star.}
    \label{fig:tess_image}
\end{figure}

There are no other TIC sources within the 42\arcsec\ $\times$ 42\arcsec\ photometric aperture (top panel of Fig. \ref{fig:tess_image}).  The brightest \gaia\ EDR3 source within 30\arcsec\ (1.5 \tess\ pixels) is 2 magnitudes fainter in \gaia\ $G$-band (a band-pass which is bluer but overlapping with the \tess\ photometer band), and the separation is 24\arcsec, thus it cannot be responsible for  dimming which is $>$16\%.  The \gaia\ Reduced Unit Weight Error (RUWE) is 1.158, consistent with a single star, and available AO data show no other sources within a few arcsec. We constructed a difference image by segregating the \tess\ images into ``in" and ``out" of occultation events, where the latter is defined by a drop in flux $>$4.15\% relative to a filtered median (see below).  This is shown in the lower panel of Fig. \ref{fig:tess_image}.  The separation in the centroid between the mean and difference images is only 0\arcsec.57, thus conclusively demonstrating that \thestar\ must be the source of the dimming signal.

\begin{figure}
	\includegraphics[width=\columnwidth]{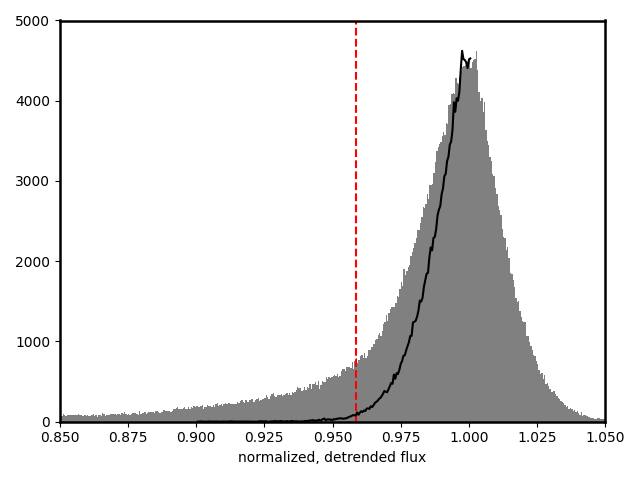}
    \caption{Distribution of Gaussian-convolved, normalized flux measurements of \thestar\ with \tess.  The black line is a reversed version of the positive side of the distribution.  The vertical red dashed line marks 4.15\% below the median, the threshold for identifying dips (see text).}  
    \label{fig:lchist}
\end{figure}

A histogram of the composite light curve is shown in Fig. \ref{fig:lchist}.  The distribution is clearly asymmetric and we identify a normalized level (95.85\%, vertical red dashed line) below which the distribution is $>$10 times that expected if it were symmetric.  About 16\% of the time the stellar brightness is below this level.  We calculated the overall quasi-periodicity parameter $Q$ and asymmetry parameter $M$ as defined by \citet{Cody2014}.  For the calculation of $M$, the overall RMS $\sigma_d$ was taken to be the outlier-resistant mean absolute deviation (MAD) of the original light curve, and for the calculation of $Q$ the systematic noise $\sigma$ was taken to be the MAD of the detrended light curve after removal of the periodic signal (see below) and exclusion of dimming events.  We found $Q=0.90$ and $M=5.4$, characteristic of a highly stochastic dipper \citep{Cody2014}.  

\begin{figure}
	\includegraphics[width=\columnwidth]{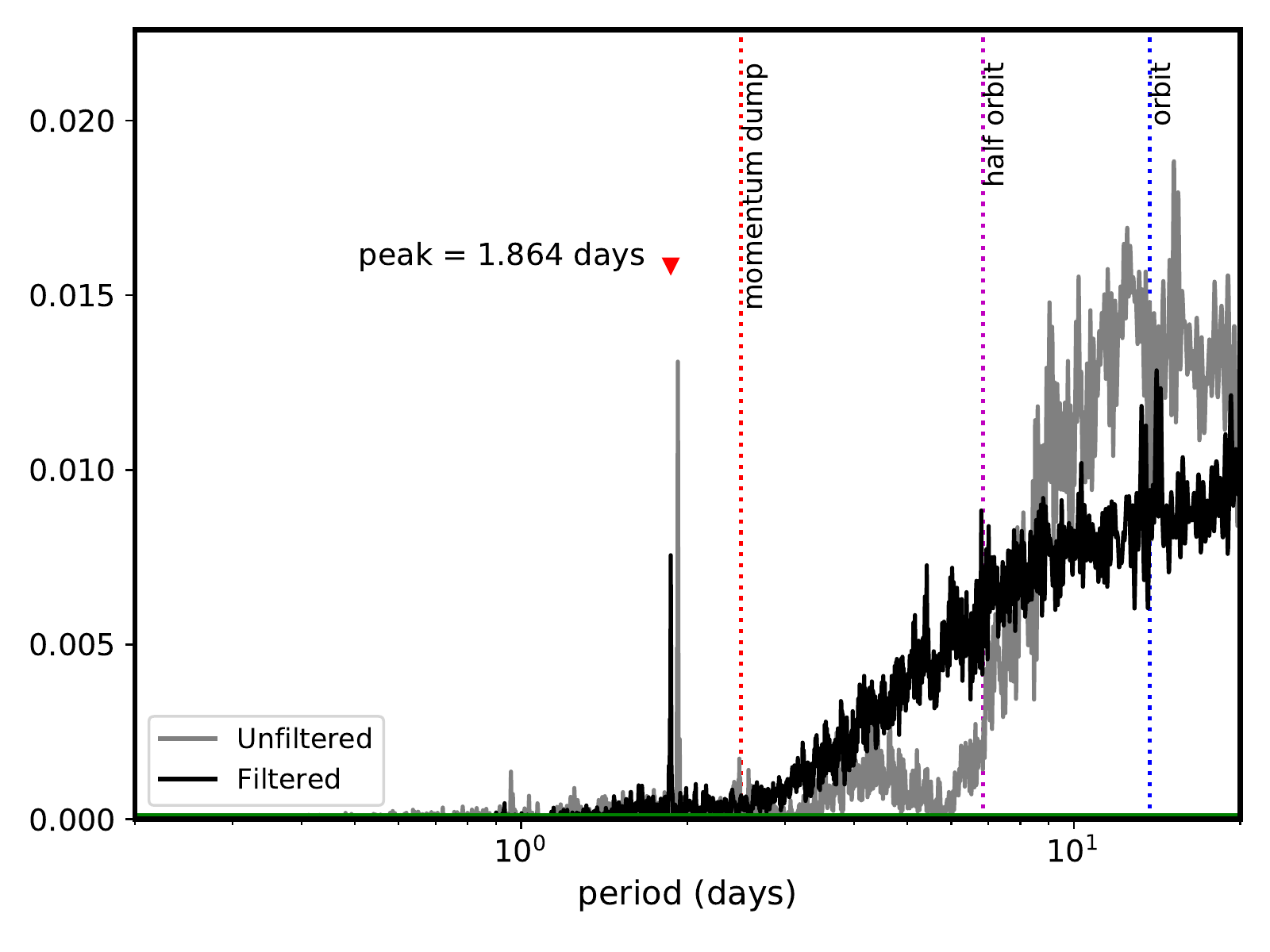}
    \caption{Lomb-Scargle periodogram of the \tess\ light curve, showing the peak at 1.863 days from the light curve where the dips have been excised (black).   These signals are clearly distinct from any artifacts produced by removal of momentum from the reaction wheels and motion of the spacecraft during an orbit (vertical dotted lines).  The black curve is the periodogram constructed after removal of dips.  The grey curve is the periodogram from the entire light curve, offset by 3\% in period for clarity.}  
    \label{fig:periodogram}
\end{figure}

\begin{figure}
	\includegraphics[width=\columnwidth]{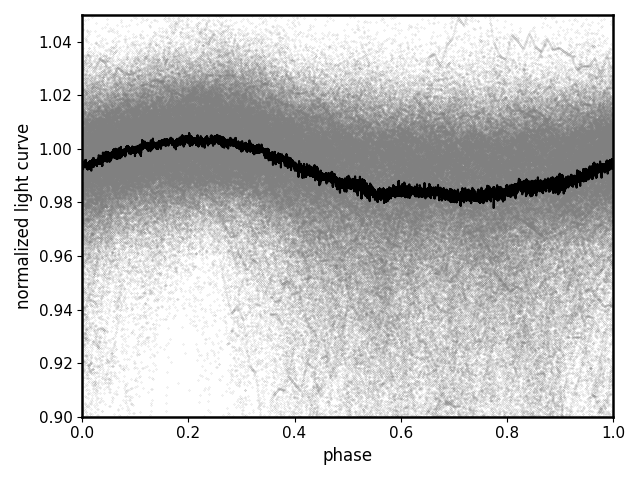}
    \caption{Normalized \tess\ light curve phased to 1.863-day period identified in the Lomb-Scargle periodogram.  The black curve is a 301-point running median filter.}  
    \label{fig:phased}
\end{figure}

We identified a pronounced $1.863 \pm 0.003$ day periodic signal in the composite light curve using a Lomb-Scargle periodogram analysis \citep[][Fig. \ref{fig:periodogram}]{Scargle1982} and determining the standard error by fitting a Gaussian to the envelope of the peak and its neighbors.  This signal is somewhat reduced but still pronounced when the analysis is repeated after excising the 16\% of the light curve in the dim state.  It is distinct from artifacts produced by removal of momentum from the spacecraft control wheels and motion of the spacecraft over an orbit (dashed lines).  This signal is much weaker ($\sim$2\% peak-to-peak) than the dimming, but is apparent when the light curve is phased to the period (Fig. \ref{fig:phased}).          

We identified dimming events in a version of the light curve smoothed by a second-order Savitzky-Golay filter with a window of 20 min.  345 candidate dimming events were identified as local minima in a 12-hour windowing of the light curve. A characteristic ``FWHM" duration for each event was calculated as the interval of time between the most recent decline below 50\% of the minimum before the minimum, and the earliest rise above 50\% of the minimum after the minimum.  We used a linearly-interpolated version of the light curve for greater time resolution.  Events with $\tau_{50}$ not exceeding twice the cadence (2 min) were discarded, leaving 329 events.  These are marked as magenta triangles in Fig. \ref{fig:tess_lc}).  

\begin{figure}
	\includegraphics[width=\columnwidth]{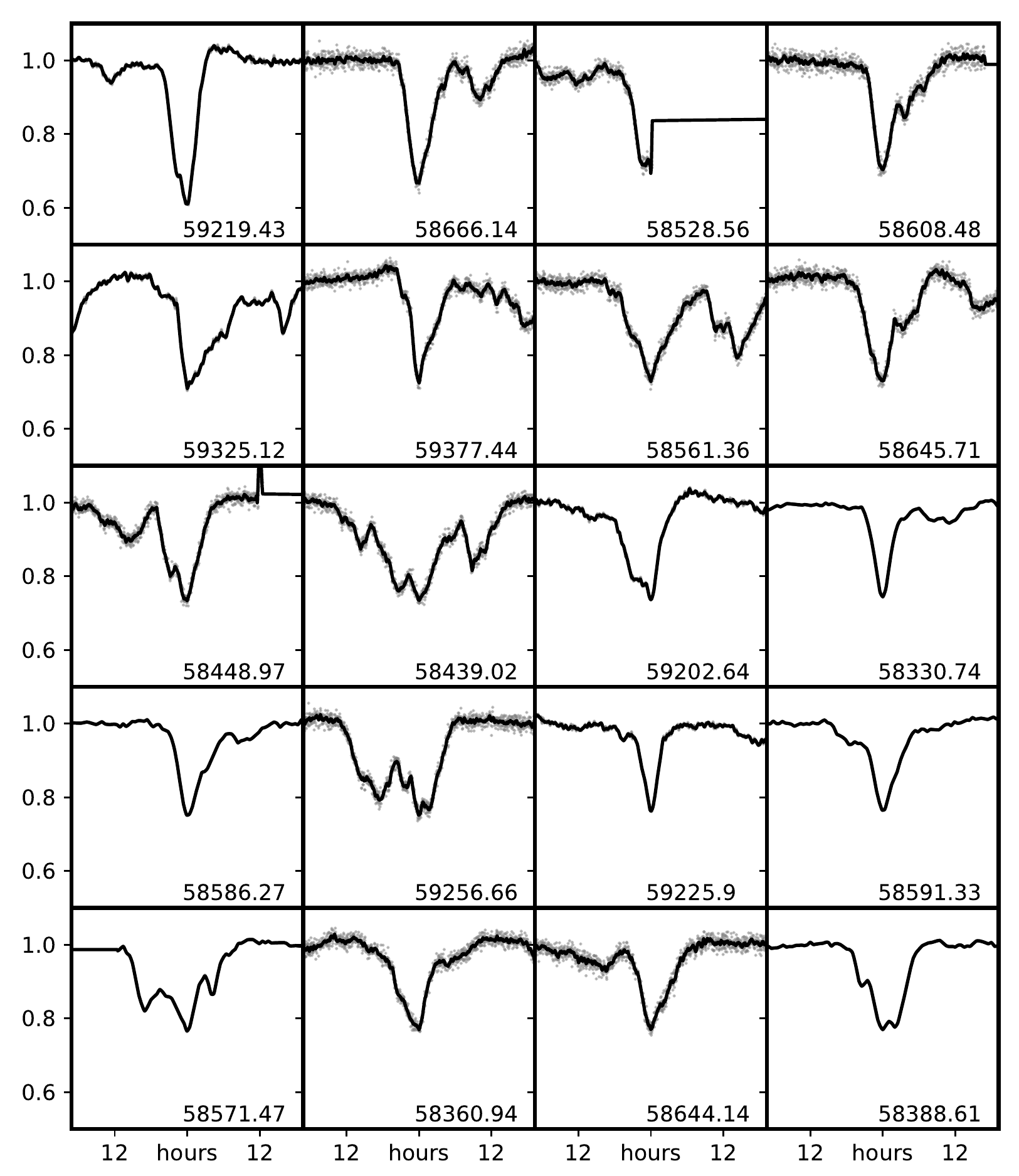}
    \caption{Normalized excerpts of \tess\ light curves of \thestar\ centered on the 20 deepest dips, arranged in decreasing order of maximum depth.  The black line is the Savitzky-Golay-filtered light curve used to identify and characterize the events.}  
    \label{fig:gallery}
\end{figure}

\begin{figure*}
	\includegraphics[width=\textwidth]{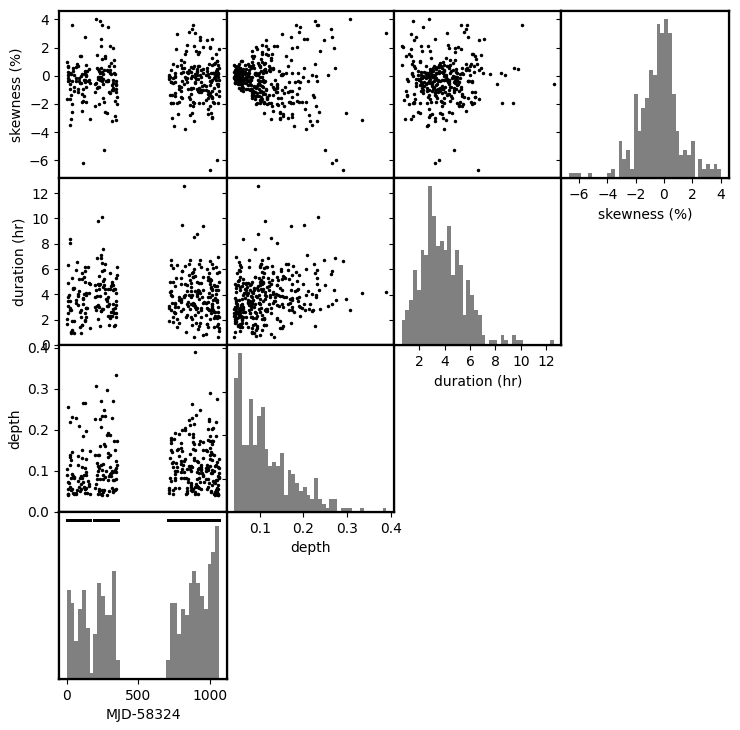}
    \caption{Triangle plot of the distribution of 329 dimming events of \thestar\ detected in the composite \tess\ light curve with time, fractional depth, duration, and skewness (Fischer's moment coefficient of skewness).  The histogram with time contains horizontal black bars indicating the observation windows.}  
    \label{fig:triangle}
\end{figure*}

Depth vs. $\tau_{50}$ as well as the time-series for each is plotted in Fig. \ref{fig:triangle}. Most events have $\tau_{50} \sim $4\,hr and nearly all have a $\tau_{50} < $6 hr.  The longest events appear to consist of clusters of 2-3  dips, while the shortest events could be artifacts of the identification process.  Figure \ref{fig:gallery} plots 1.6-day subsets of the \tess\ light curve centered on the twenty deepest dips.  We computed the skewness in time of each event using Fischer's coefficient of skewness:
\begin{equation}
    \label{eqn:skewness}
    \Tilde{\mu}_3 = \frac{\int_{t_1}^{t_2} dt f(t) \cdot \left(t-\bar{t}\right)^3}{\left[\int_{t_1}^{t_2} dt f(t) \cdot \left(t-\bar{t}\right)^2\right]^{3/2} \left[\int_{t_1}^{t_2} dt f(t) \right]^{1/2}},
\end{equation}
where $\bar{t} = \left(\int_{t_1}^{t_2} dt f(t) \cdot t\right) /\left( \int_{t_1}^{t_2} dt f(t)\right)$ and the integrals are over the event's $\tau_{50}$ duration.  The distribution is plotted in Fig. \ref{fig:triangle}; $\Tilde{\mu}_3$ is typically small ($\lesssim5$\%), but there is a significant excess in negative values (195 compared to 134) such that there are more dips with leading tails. (We also calculated skewness over the 30\% of dip depth range and obtained similar results.)

\begin{figure}
	\includegraphics[width=\columnwidth]{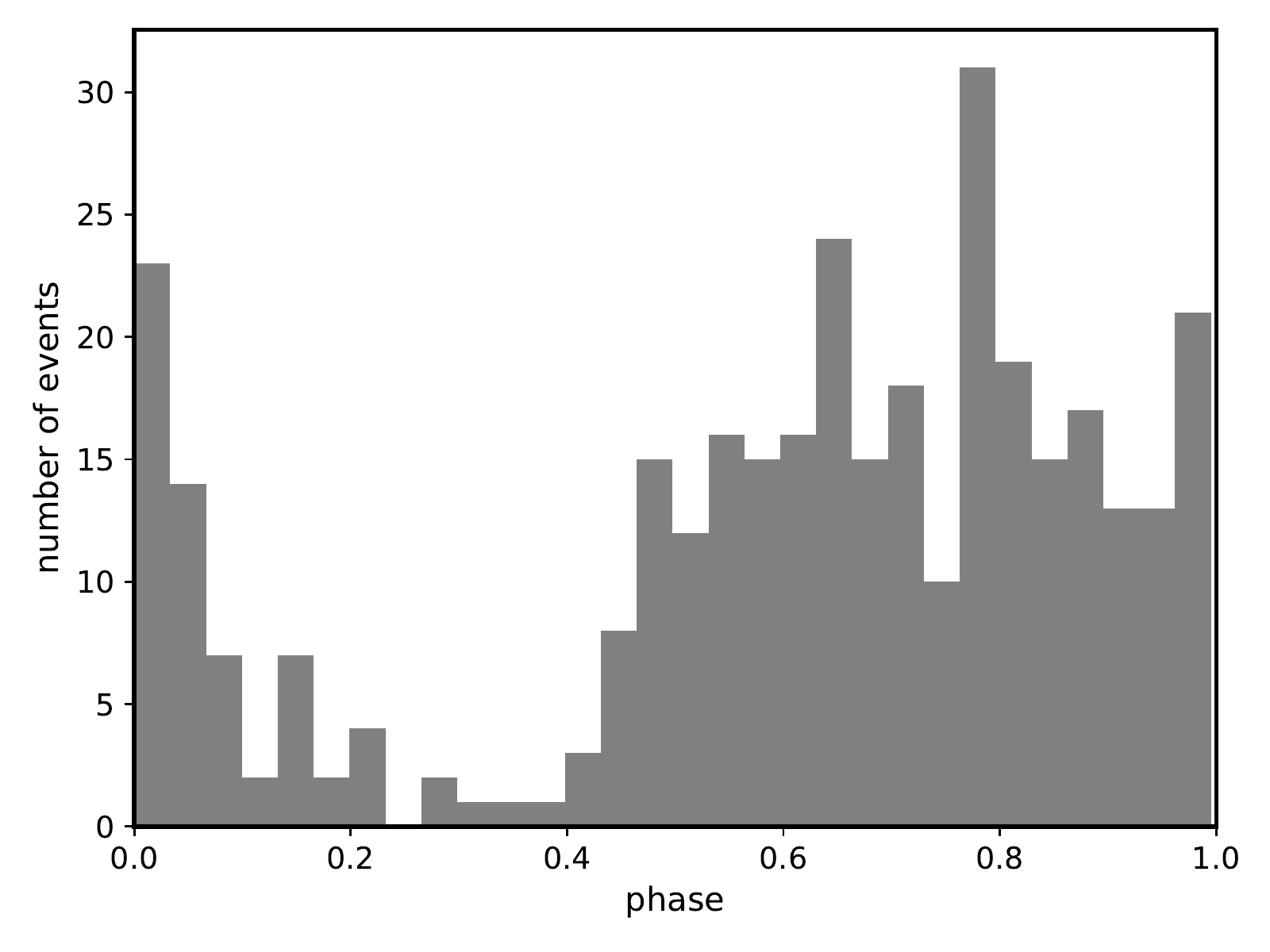}
    \caption{Distribution of dimming events with phase of the 1.863-day periodic signal, which is possibly variability from the rotation of the star.  The phase here is the same as in Fig. \ref{fig:phased}.}  
    \label{fig:dip_phase}
\end{figure}

The distribution of dimming events with the phase of the periodic 1.863-day signal identified in our periodogram analysis is non-uniform (Fig. \ref{fig:dip_phase}).  A Kolmorogov-Smirnov test compared to a uniform distribution returns $D=0.25$ and $p=3.4 \times 10^{-19}$.  This is also apparent in the phased light curve (Fig. \ref{fig:phased}), thus it is not simply a bias to the additive effect of stellar rotational variability and dimming.  Dips are more likely to occur in the dimmer phase of the star's rotational light curve, perhaps when a more spotted region of the star faces us.

\begin{figure}
	\includegraphics[width=\columnwidth]{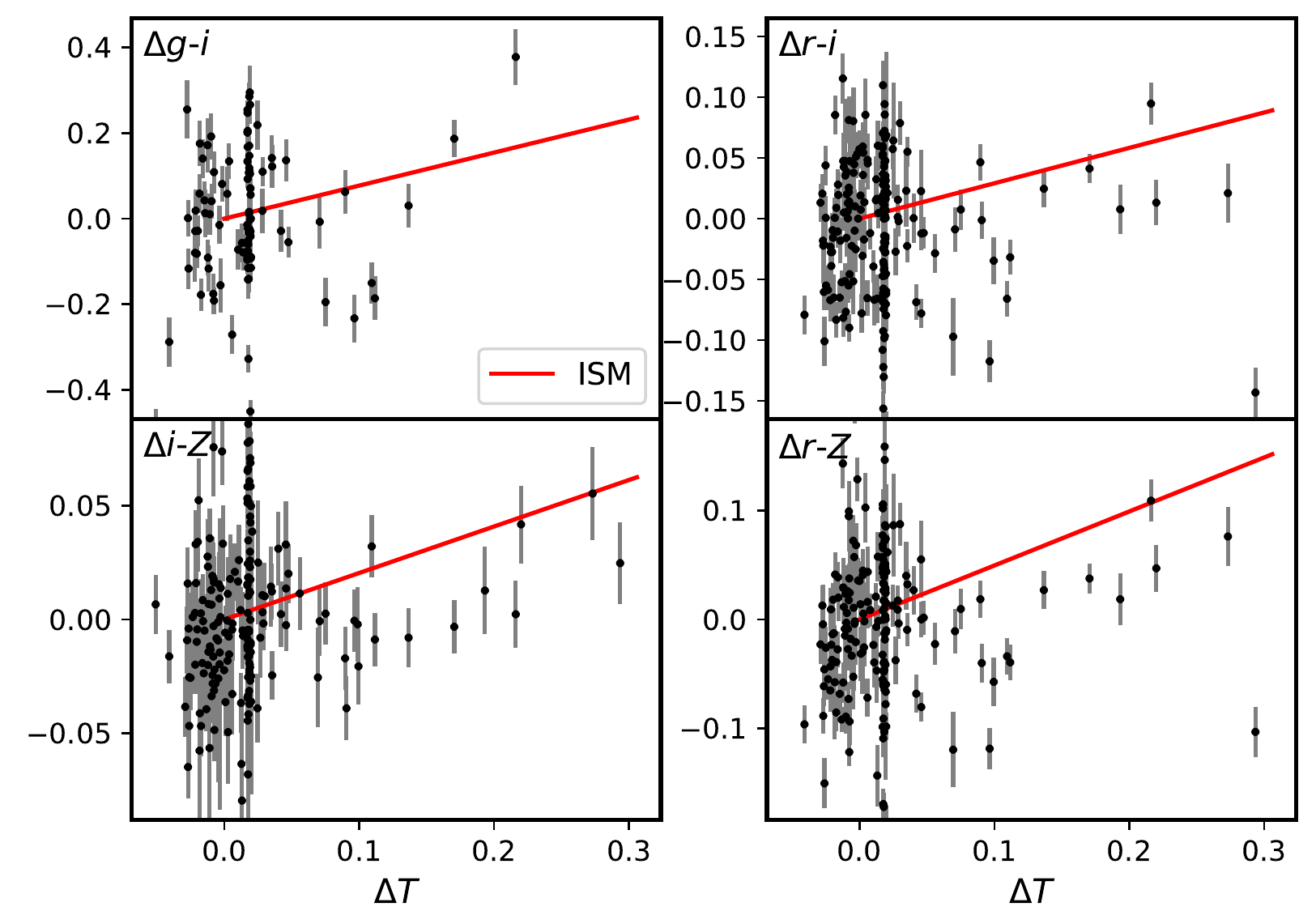}
    \caption{Extinction-reddening plot for combinations of the Sloan $gri$ and Pan-STARRS $Z$ photometry from LCOGT observations of \thestar.  The red lines are the trends expected for ISM-like dust using the extinction coefficient for the LCOGT filters from \citet{Yuan2013} and for the \tess\ $T$-band from \citet{Stassun2018}.  ``Gray'' extinction by large particles would be a horizontal trend in these plots.}  
    \label{fig:extinct-redden}
\end{figure}

The multi-bandpass photometry of LCO ($griZ$, spanning 3800-9350\AA) can measure reddening during dimming and constrain the grain-size distribution and composition of the occulting dust \citep{Budaj2015}.  We used the \tess\ data to identify 38 dimming events during the 50-day interval that LCO and \tess\ were both observing \thestar\ (Fig. \ref{fig:lco}).  Reddening-extinction diagrams were constructed by comparing variation in LCOGT-based colors to changes in \tess\ $T$ magnitudes.  We  corrected for the variability of the star over the finite time each set of multi-bandpass observations took place (up to an hour) by subtracting the known variation in the \tess\ magnitude over this interval multiplied by the ratio of extinction coefficients $(R_1 - R_2)/R_T$ where the indices 1 and 2 refer to the two LCOGT band-passes used to construct the color, and $T$ refers to \tess.  We assumed ISM values for extinction coefficients (see below) but because the variation in $\Delta T$ is small and departures from this assumption would have a minor effect.  The reddening-extinction diagrams (Fig. \ref{fig:extinct-redden}) for the $riZ$ pass-bands show clear trends that are consistent or even slightly steeper than expected for ISM-like dust (red lines).  Significant scatter in $r$-band photometry is expected from variable stellar H$\alpha$ emission (Fig. \ref{fig:soar}) and in $Z$-band photometry due to detector fringing.  Elevated extinction in the $g$ band is strongly suggested in LCOGT data obtained non-contemporaneously with \tess\ (top panel of Fig. \ref{fig:lco}) and there is a possible trend in $\Delta g-i$ vs. $\Delta T$ in Fig. \ref{fig:extinct-redden}, but the poor quality of the $g$-band photometry due to low SNR and poor tracking preclude any definitive conclusion.

\subsection{Stellar properties}
\label{sec:parameters}

\begin{table}
\begin{center}
\caption{Properties of \thestar} 
\label{tab:params}
\begin{tabular}{l | l | l}
\multicolumn{1}{c}{\textbf{Catalog}} & \multicolumn{1}{c}{\textbf{ID}} &  \\
\hline
Gaia EDR3$^{a}$ & 5280226578890455168 &  \\
TIC$^{b}$ & 167303776 & \\
2MASS PSC$^{c}$ & J06320799-6810419	& \\
AllWISE$^{d}$ & J063207.99-681041.6 & \\
UCAC4$^{e}$ & 110-012461 &  \\
\hline
\multicolumn{1}{c}{\textbf{Property}} & \multicolumn{1}{c}{\textbf{Value (error)}} & \multicolumn{1}{c}{\textbf{Source}}\\
\hline
Right Ascension $\alpha$ & 06h 32m 07s.97 & \multirow{6}{*}{\gaia\ EDR3$^{a}$}\\
Declination $\delta$ & -68d 10\arcmin\ 41\arcsec.99 & \\
$\mu_{\alpha}$ $\alpha$ [mas yr$^{-1}$]& 9.350 (0.033) & \\
$\mu_{\delta}$ [mas yr$^{-1}$]& 33.936 (0.034) & \\
distance [pc] & 93.0 (0.2) & \\
RUWE & 1.158 & \\
RV [km s$^{-1}$] & +22.5 (2.3) & this work\\
\midrule
$m_{\rm FUV}$ & 21.49 (0.35) & \multirow{2}{*}{\emph{Galex} GR6+7$^{f}$} \\
$m_{\rm NUV}$ & 20.66 (0.21) & \\
\midrule
$B_p$ & 17.297 (0.012) & \multirow{3}{*}{\gaia\ EDR3$^{a}$}\\
$R_p$ & 14.240 (0.005) & \\
$G$ & 15.518 (0.003) & \\
\midrule
$I$ & 14.07 (0.03) & \multirow{3}{*}{DENIS DR3$^{g}$}\\
$J$ & 12.41 (0.06) & \\
$K_s$ & 11.43 (0.08) & \\
\midrule
$J$ & 12.395 (0.026) & \multirow{3}{*}{2MASS PSC$^{c}$}\\
$H$ & 12.798 (0.027) & \\
$K_S$ & 11.50 (0.023) & \\
\midrule
W1 [3.4] & 11.258 (0.022) & \multirow{4}{*}{AllWISE$^{d}$}\\
W2 [4.6] & 10.96 (0.02) & \\
W3 [12] & 9.195 (0.022) & \\
W4 [25] & 7.519 (0.065) & \\
\hline
\multicolumn{3}{c}{\textbf{Inferred properties}}\\
\hline
\teff\ [K] & 3100 (75) & SED fit\\
SpT & M4.5 (0.5) & photometry, \teff\\
$L_\star$ [\lsun] & $1.324 (0.008) \times 10^{-2}$ & SED + parallax\\
$R_\star$ [$R_{\odot}$] & 0.40 (0.02) & $L_\star$ and teff\ \\
log~g & 4.0 (0.5) & SED fit\\
$M_\star$ [$M_{\odot}$] & 0.13-0.23 & isochrone fit\\
age [Myr] & 30-60 & isochrone fit + Li \\
rotation period [days] & 1.863 (0.003) & \tess\ photometry\\
\hline
\end{tabular}
\end{center}
$^a$\citet{Gaia2021}. $^{b}$\citet{Stassun2018}. $^{c}$\citet{Skrutskie2006}. $^{d}$\citet{Cutri2013}. $^{e}$\citet{Zacharias2013}.  $^{f}$\citet{Bianchi2017}.  $^{g}$\citet{DENIS2005}.
\end{table}

Our AO imaging rules out near equal-brightness companions between 0\arcsec.05 and 0\arcsec.7 (Sec. \ref{sec:ao}).  \gaia\ can resolve more widely-separated binaries and a search of EDR3 within 3 deg. of \thestar\ identified one star (5280267398259251712) 34\arcmin.6 away at about the same distance (89.8 $\pm$ 1.4 pc) as \thestar\ (93.0 $\pm 0.2$) and with a similar but statistically different proper motion ($\mu_{\alpha}=10.879\pm0.237$, $\mu_{\delta}=33.580\pm0.205$ mas yr$^{-1}$).  With $M_G=12.1$ and $B_p-R_p$ the star falls close to the best-fit isochrones for the Carina moving group (see Sec. \ref{sec:age}) and is likely a fellow member, but the projected separation is nearly $2\times10^5$~au and the two stars cannot be bound.  The Reduced Unit Weight Error of \thestar, a measure of the goodness-of-fit of a single-star solution to the astrometric data, is 1.158, consistent with a single star \citep{Belokurov2020}.  We therefore assumed that \thestar\ is a single star in our subsequent analysis.

Photometry of \thestar\ was obtained from \gaia\ \citep[$G$,$B_P$,$R_P$,][]{Riello2021}, the Deep Near Infrared Survey of the Southern Sky \citep[DENIS $I$ and $K_s$ ,][]{Schuller2003,DENIS2005}, VISTA $J$ and $K_s$ \citep{Cross2012}, 2MASS $JHK_s$ \citep{Skrutskie2006}, and \wise\ 3.4, 4,4, 12, and 25 $\mu$m \citep{Wright2010,Cutri2013}.   Based on the 3-d dust map of \citet{Leike2020} (see Sec. \ref{sec:age} for details), we estimated interstellar extinction of $A_V = 0.03$.  The extinction-corrected spectral energy distribution (SED, Fig. \ref{fig:sed}) was fit to stellar models using the SED analysis tools of the Virtual Observatory \citep{Bayo2008}.  The VISTA photometry was ultimately excluded from the fit as redundant, with anomalously small errors, and the AllWISE data were excluded due to excess emission (see Sec. \ref{sec:disk}).  We assumed a near-solar metallicity for \thestar\ due to its young age.  The best-fit ($\chi^2=$28.4 with $\nu = 10$ degrees of freedom) solar-metallicity BT-SETTL model with \citet{Caffau2011} relative abundances has \teff=3100K and $\log g$=4.  (Stellar surface gravity is very poorly constrained by photometry and should be regarded as merely a fitting parameter.)  As a check, we compared our $JHK$ spectrum (Fig. \ref{fig:tspec}), particularly the CO lines in $K$-band, to PHOENIX model spectra using the {\tt STARFISH} emulator \citep{Czekala2015} and found good agreement with \teff=3100K, $\log g$=4, and [Fe/H]=0. A PHOENIX model spectrum with these parameters from \citet{Husser2013} is plotted in Fig. \ref{fig:sed}.  \citet{Pecaut2013} assign a spectral type of about M4.5-5 to a \teff\ of 3100K.

Stellar luminosity was calculated by combining the integrated best-fit SED and the \gaia\ parallax. The radius was determined using the luminosity, temperature, parallax, and the Stefan-Boltzmann equation.  Masses were found by applying Eqn. \ref{eqn:mass} to the  best-fit isochrone in each set of models compared to the stellar parameters of Carina moving group stars (see Sec. \ref{sec:age}).  The different model sets yielded different masses (inversely related to age) and these ranged between 0.13 and 0.23\msun.  For calculations involving the stellar mass, we adopted 0.18\msun.  Note that this implies a slightly higher gravity of $\log$\,g of 4.5, but still consistent with that expected of a pre-main sequence M dwarf.

\begin{figure}
	\includegraphics[width=\columnwidth]{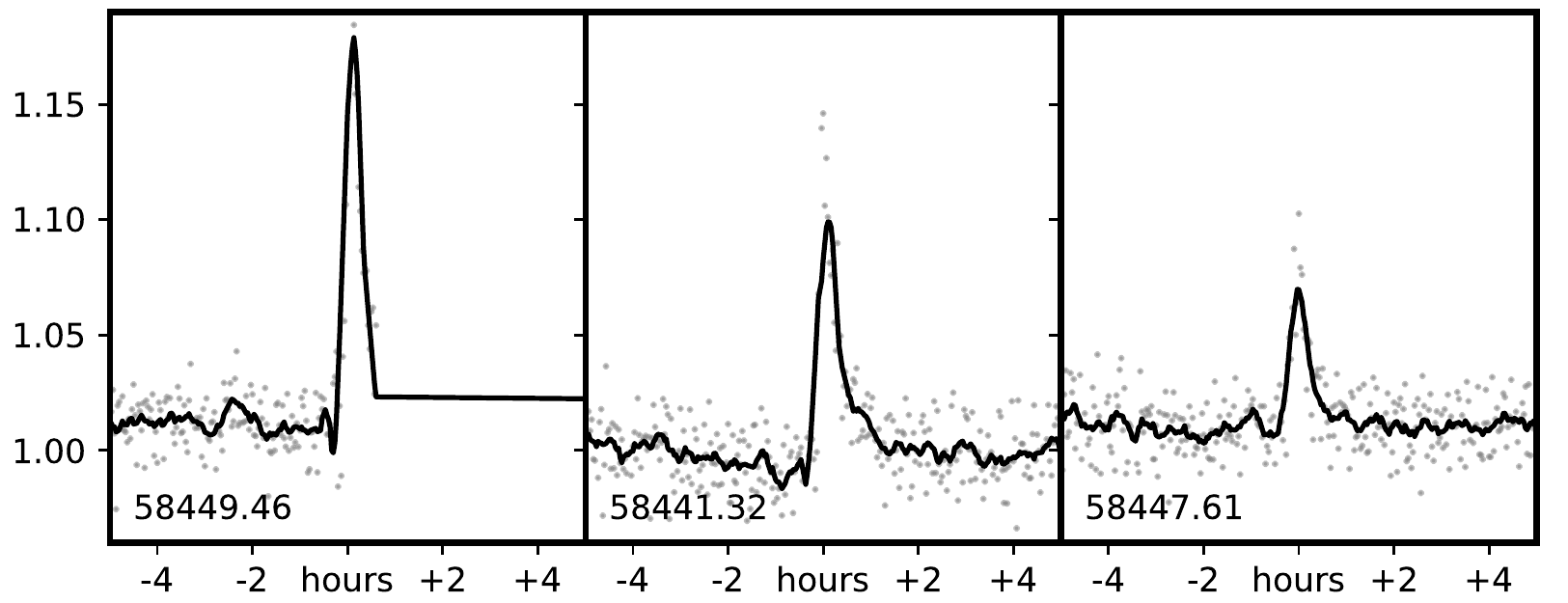}
    \caption{Large flares identified in the \tess\ light curve of \thestar. The black line is the Savitzky-Golay-filtered light curve used to identify and characterize the events.}  
    \label{fig:flares}
\end{figure}

The rapid rotation ($P=1.86$ days, Sec. \ref{sec:tess_analysis}) of \thestar\ is an indicator of a young age for this single star.  Elevated emission in the Balmer $\alpha$ line of H\,I and the ultraviolet are also indicators, but in this case the signals seem dominated by accretion (Sec. \ref{sec:disk}).  \thestar\ is flaring, but only modestly so; we detected just three large flares (amplitudes greater than 5\% quiescent level) in the entire light curve, all within an interval of about 8 days time (Fig. \ref{fig:flares}). This could be the result of the appearance of an especially active region on the star.  

\subsection{Circumstellar Disk}
\label{sec:disk}

Our optical and infrared spectra of \thestar\ contain multiple signatures of an accreting disk, including lines of H~I, [OI], and triplet He~I (Figs. \ref{fig:soar} and \ref{fig:tspec}).  The equivalent width (EW) of the H$\alpha$ line was calculated using the line and pseudo-continuum regions defined in \citet{Lepine2013} and determined to be 22.1\AA.  The 10\% kinematic width $\nu_{10}$ was calculated as 336 \kms.  Both of these measures are well in excess of established criteria for distinguishing accreting vs. non-accreting (and merely magnetically active) stars \citep{White2003,Mohanty2005}, and can be used to estimate accretion rate.  The $\dot{M}$-\halpha\ EW relation of \citet{Herczeg2008} yields an accretion rate of $6 \times 10^{-12}$ \msun$^{-1}$, while the $\dot{M}-\nu_{10}$ relation of \citet{Natta2004} gives a significantly higher accretion rate of $2 \times 10^{-10}$ \msun$^{-1}$, with an uncertainty of a factor of $\sim$2.  

The EW of the optical triplet He\,I line was estimated over the range 6676.8-6680.8\AA\ to be 0.71\AA.  Unfortunately, our optical spectrum is not flux-calibrated nor does it completely span the wavelength range of any filter with available photometry for \thestar.  Instead, we use the model spectrum from \citet{Husser2013} to compute the continuum flux density at 6678\AA\ relative to the integrated flux over the \gaia\ $Rp$ pass-band.  We arrive at a line flux of $7.3 \times 10^{-16}$ ergs sec$^{-1}$ cm$^{-2}$ which at the distance of the star is a luminosity of $7.6\times 10^{26}$ ergs sec$^{-1}$ or $2 \times 10^{-7}$\lsun. Using the scaling relations between indicator line luminosity and accretion luminosity of \citet{Herczeg2008} we estimated the latter to be $1.0 \times 10^{-4}$\lsun. Using Eqn. 8 in \citet{Gullbring1998} with the disk inner edge at 5$R_\star$, we estimate an accretion rate of $8 \times 10^{-12}$ \msun\ yr$^{-1}$, in good agreement with the \halpha\ EW-based estimate, but not the \halpha\ 10\% width-based estimate.  We note the latter is affected by stellar inclination \citep{Curran2011}, and that the \citet{Natta2004} relation was developed for objects 0.01-0.1\msun, i.e. somewhat less massive than \thestar.  We have not corrected for continuum veiling, but this is expected to be negligible at these low accretion rates.

The profile of the He\,I triplet line at 1.085 \micron\ contains blue-shifted ($\approx$-20 \kms) emission and red-shifted ($\approx$+100 \kms) absorption that is characteristic of many T Tauri stars \citep[Fig. \ref{fig:tspec},][]{Kwan2007}.  Modestly blue-shifted emission arises from a hot wind moving nearly perpendicular to the line of sight, plus red-shifted sub-continuum absorption is the hallmark of ``funneled" accretion flows.  Although there is no indicator of blue-shifted absorption due to an intervening wind, the limited spectral resolution cannot preclude some weak absorption.

\thestar\ is also excessively luminous in the far-ultraviolet (FUV, 1340-1806\AA) and near-ultraviolet (NUV, 1693-3007\AA) channels of \galex\ \citep{Bianchi2017} relative to expectations for its photosphere.  But its UV luminosity is low (Table \ref{tab:params}), compared to YSOs in the Taurus-Aurigae star-forming region \citep{GomezdeCastro2015} and at the detection limits of \galex.   \thestar\ has a FUV-NUV color of 0.8, similar to that of classical T Tauri stars, disfavoring interstellar or circumstellar extinction as the primary cause of its UV faintness  \citep{GomezdeCastro2015}.  Instead, its low accretion rate may be responsible (Sec. \ref{sec:disk}).

\begin{figure}
	\includegraphics[width=\columnwidth]{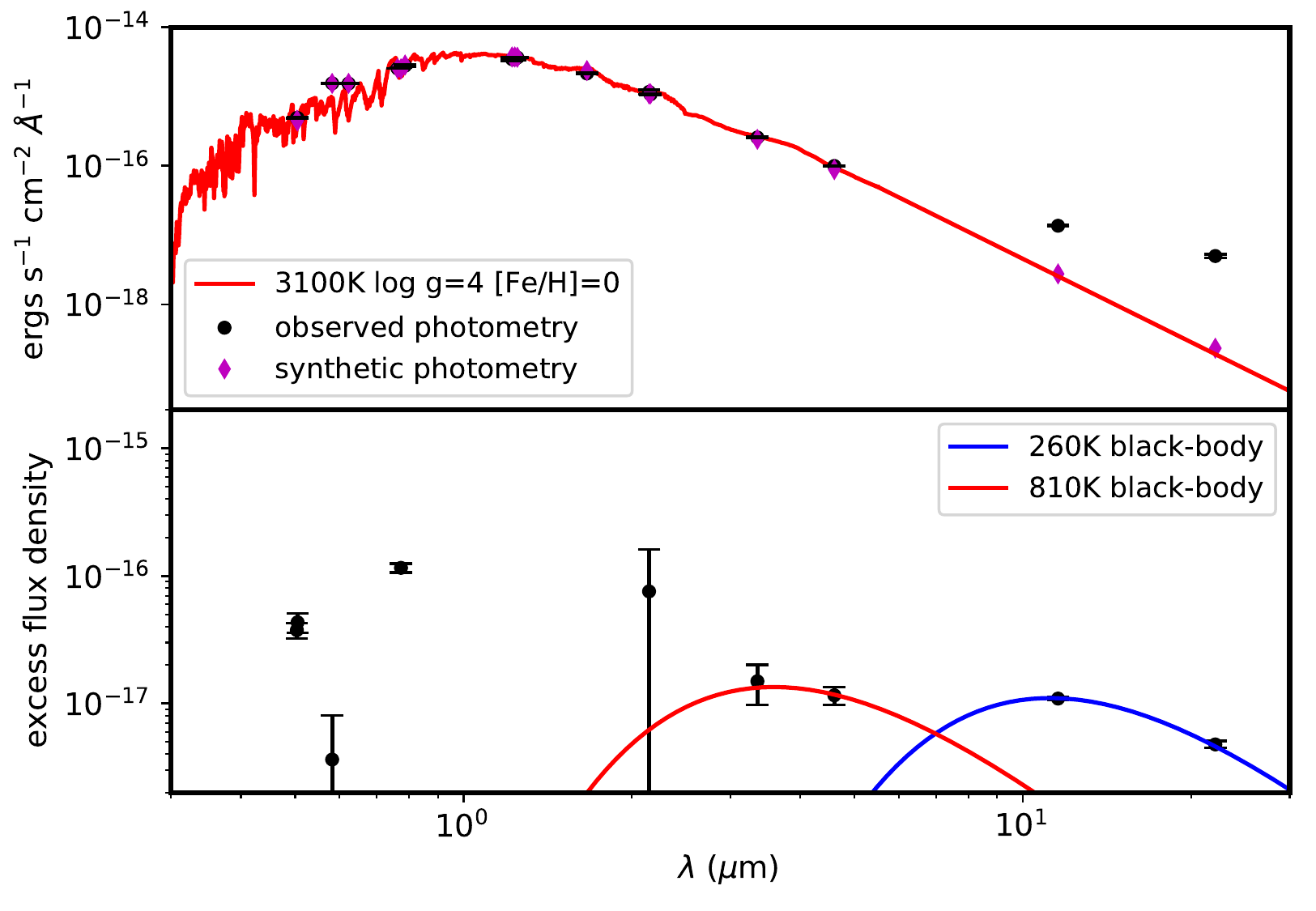}
    \caption{Top: Spectral energy distribution of \thestar\ using \emph{Gaia},DENIS, 2MASS, and \emph{WISE} photometry.  Black points are measurements, and magenta points are synthetic photometry produced from the best-fit {\tt PHOENIX} stellar model.  The solid red line is a {\tt PHOENIX} model atmosphere with parameters \teff=3100K, $\log g = 4.0$, [Fe/H] = 0, [$\alpha$/Fe] = 0.  Bottom: difference between the SED and stellar atmosphere model, showing the excess emission at $\lambda > 2$\micron compared to two black-body SEDs.}
    \label{fig:sed}
\end{figure}

The disk of \thestar\ also manifests itself as excess emission at the longest wavelengths. There is a pronounced excess in the 12- and 25-\micron\ channels of \wise, and smaller but still formally significant excesses in the 3.4 and 4.6-\micron\ channels (Fig. \ref{fig:sed}).  Minimum $\chi^2$ fitting shows that no single black-body can explain this emission, suggesting that instead there is a disk or at least multiple belts of circumstellar material.   

\begin{figure}
    \centering
    \includegraphics[width=\columnwidth]{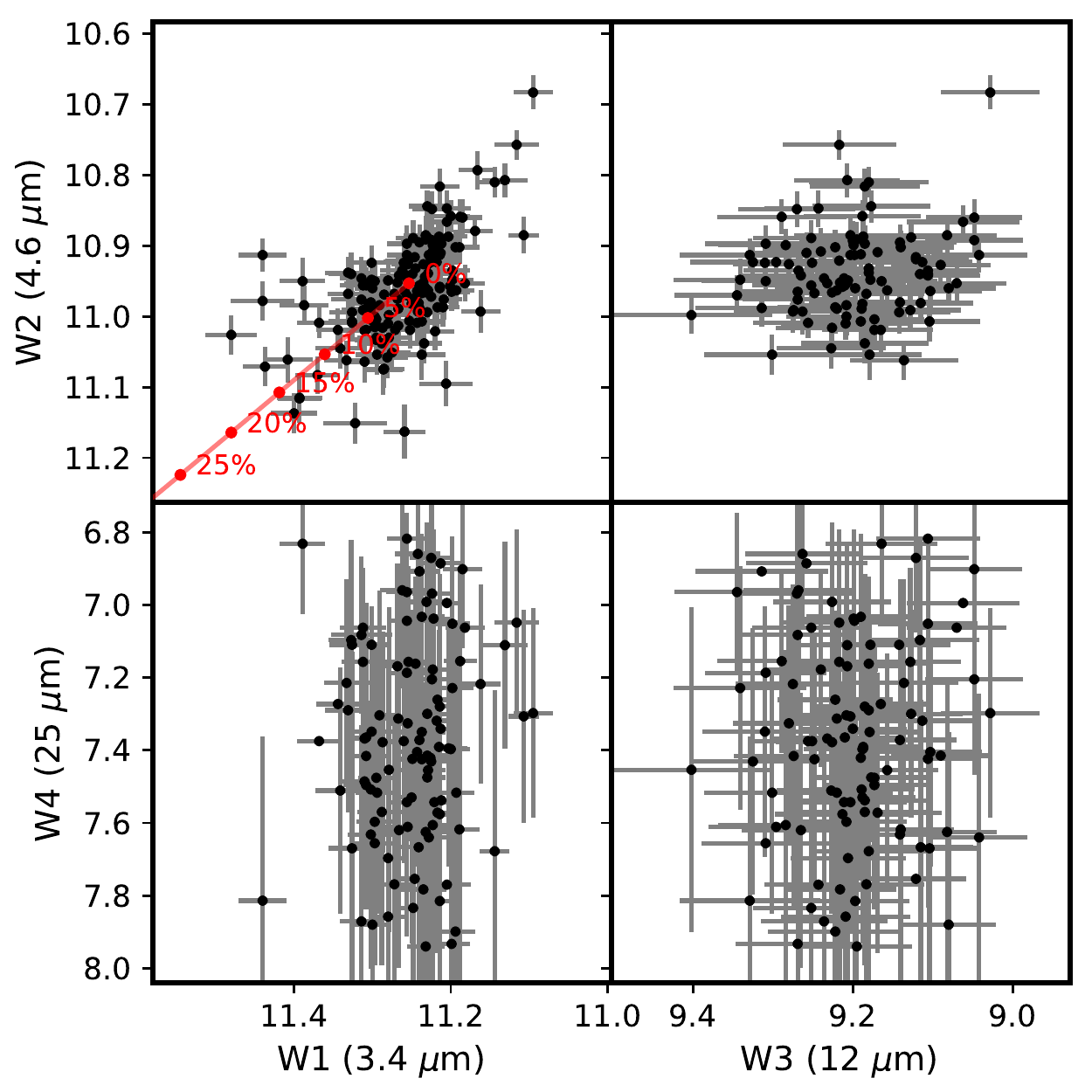}
    \caption{ALLWISE multi-epoch photometry showing significant, positively correlated variability at 3.4 and 4.6 \micron, but not at longer wavelengths.  The red line in the upper left panel is the variability predicted for achromatic extinction of the stellar photosphere, with \% increments of extinction marked.}
    \label{fig:wise_multi}
\end{figure}

AllWISE multi-epoch photometry of \thestar\ show significant variability in the \wise\ W1 and W2 channels ($\chi^2$ of 862 and 1180, respectively, for $\nu$=160 degrees of freedom; Fig. \ref{fig:wise_multi}).  The W1 and W2 emission is significantly correlated (Spearman non-parametric rank test $p=1 \times 10^{-18}$).   In contrast, there is no detected variability in the longer-wavelength (W3 and W4 channels ($\chi^2$ of 113 and 58 for 106 degrees of freedom, respectively, for a fit to a constant value).  Some of the variability in W1 and W2 could be due to variable extinction along the line of sight occulting the star (but not the cooler disk) but only if the extinction is largely achromatic, i.e., the occulting material has a large grain size (Fig. \ref{fig:wise_multi}).  This would be in conflict with the LCO results (Fig. \ref{fig:extinct-redden}).  Also, it appears that the outer regions of the disk that dominate long-wave emission are not variably shadowed by inner disk material \citep[``see-saw" variability,][]{Muzerolle2009}.  

\subsection{Moving Group Membership and Age}
\label{sec:age}

\begin{figure*}
	\includegraphics[width=0.495\textwidth]{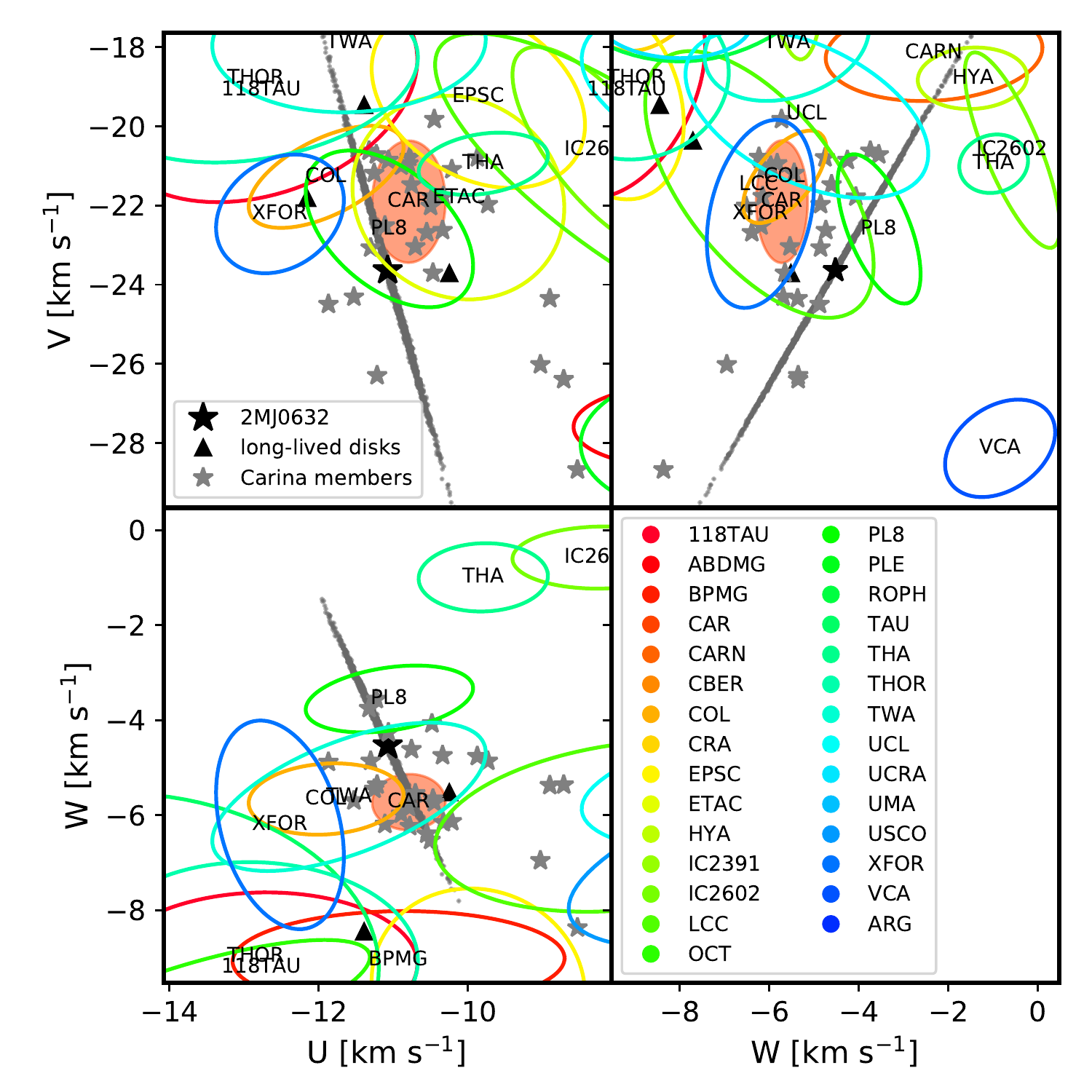}
		\includegraphics[width=0.495\textwidth]{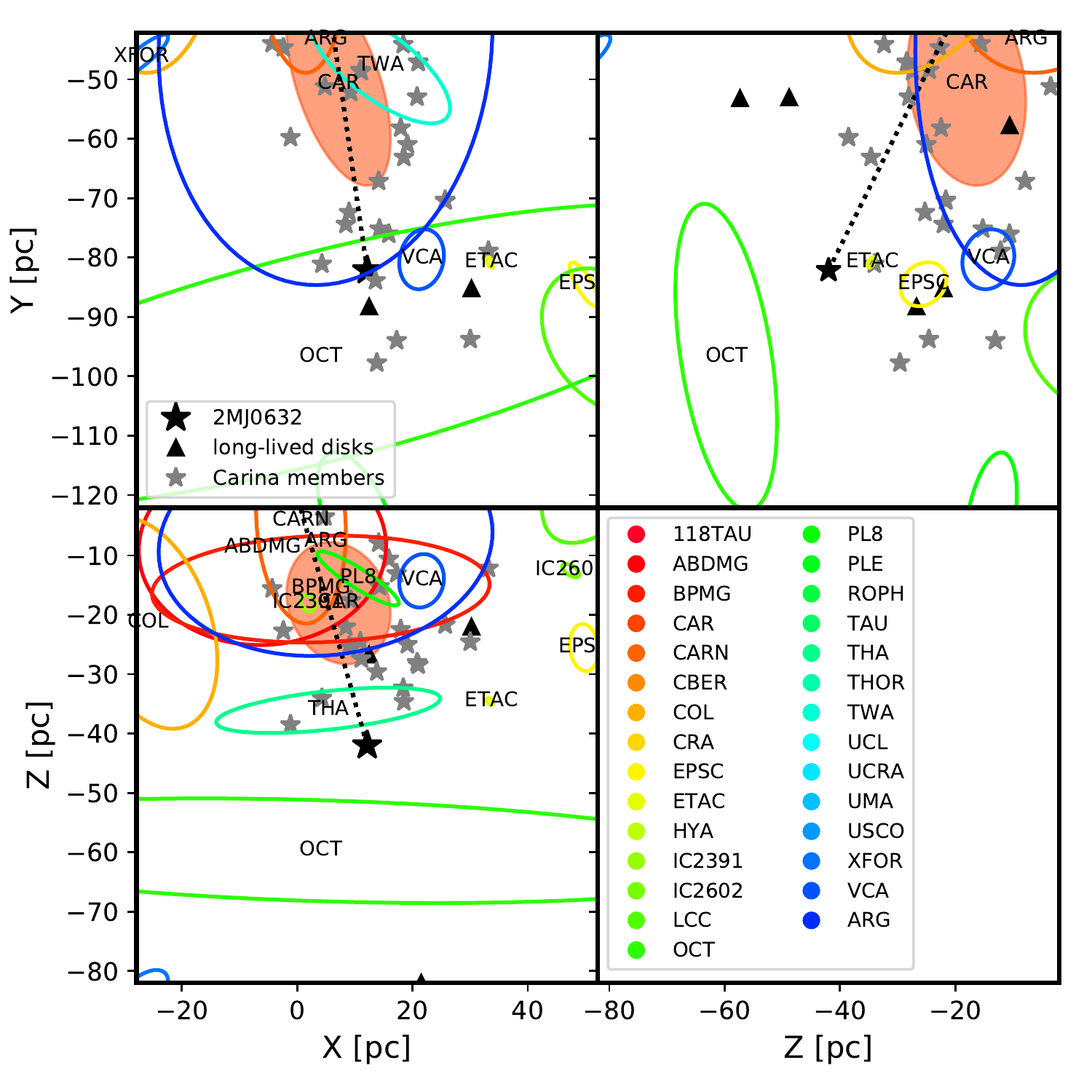}
    \caption{\textbf{Left:} $UVW$ Galactic space motion of \thestar\ (black star) compared to those of nearby young moving groups, clusters, and star-forming regions in the {\tt Banyan $\Sigma$} database  \citep{Gagne2018}.  The size of the ellipses represent the 1-$\sigma$ velocity dispersion of the groups.  The Carina moving group is the filled ellipse.  The uncertainty in the space motion of \thestar, dominated by the error in the RV, is represented by the distribution of grey points.  Other members of the Carina moving group with Banyan membership probabilities $>$90\% from \citet{Booth2021} are plotted as grey stars. Several other stars with long-lived disks \citep{Silverberg2020} are plotted as black triangles. \textbf{Right:} Left-handed $XYZ$ Galactic coordinates of the \thestar\ (black star) compared to those of nearby young moving groups, clusters, and star-forming regions in the {\tt Banyan $\Sigma$} database  \citep{Gagne2018}.  The size of the ellipses represent the 1-$\sigma$ spatial dispersion of the groups.  The Carina moving group is the filled ellipse.  Other members of the Carina moving group with Banyan membership probabilities $>$90\% from \citet{Booth2021} are plotted as grey stars. Several other stars with long-lived disks \citep{Silverberg2020} are plotted as black triangles.  The dashed line is the line-of-sight to \thestar. }  
    \label{fig:uvwxyz}
\end{figure*}

Based on \gaia\ EDR3 astrometry and our measured RV, we calculated Galactic space motions relative to solar of $(U,V,W) = (-11.08 \pm 0.29,-23.6 \pm 2.0, -4.5 \pm 1.0)$ km~sec$^{-1}$.  Applying the Bayesian {\tt Banyan $\Sigma$} algorithm \citep{Gagne2018} to the 6-dimensional space $XYZ$ and motion $UVW$ coordinates, we found a 92.3\% probability of membership in the Carina moving group and negligible probability for every other cluster and moving group in the database; the remainder probability is assigned to the field.  The $UVW$ posterior distribution of \thestar\ passes through the cluster of Carina members cataloged by \citet{Booth2021} and intercepts the center of the moving group (Fig. \ref{fig:uvwxyz}).  \thestar\ lies in the neighborhood of the Octans group (Fig. \ref{fig:uvwxyz}) but is kinematically distinct.  It also passes close to the Columba moving group and the Platais 8 open cluster, but the star is distant from these two in spatial coordinates $XYZ$ (Fig. \ref{fig:uvwxyz}).  However, \citet{Torres2008} previously related Carina with Columba and the Tucana-Horologium groups, all of which appear co-eval \citep{Bell2015}.  \citet{Gagne2021} proposed that Carina, Columba, and the Theia 92, 113, and 208 moving groups are all derived from the Platais 8 open cluster \citep{Platais1998}, possibly by partial disruption of the cluster into ``tails", and are all thus co-eval.      

Based on a comparison with PARSEC, Dartmouth, Pisa and \citet{Baraffe2015} isochrones, \citet{Bell2015} estimated the age of the Carina moving group as about 45 Myr.  In contrast, \citet{Schneider2019} used Li abundances to estimate an age similar to that of the Beta Pictoris moving group (20-25 Myr).  More recently, \citet{Booth2021} made an isochrone-based estimate of $\sim$15 Myr.    We revisited the question of age with the larger sample of \citet{Booth2021}, incorporating improved astrometry and photometry from \gaia\ EDR3 \citep{Gaia2021}.  \gaia\ EDR3 also allowed us to reject additional binaries based on a high value of the Renormalized Unit Weight Error (RUWE); we excluded stars with RUWE $>1.4$ as very likely binaries, stars bluer than $B_p-R_p<0.75$ which are not informative for age estimation, and both \thestar\ and WISE J0808 because variable extinction might affect the $G$-band \citep{Silverberg2020}.

Carina members lie beyond the dust-free Local Bubble and we corrected \gaia\ astrometry for extinction using the 3-d dust maps of \citet{Leike2020} as implemented with the {\tt DUSTMAPS} routine \citep{Green2018}.  The \citet{Leike2020} dust maps estimate extinction density in the \gaia\ $G$ pass-band which we converted to $A_V$ and then to $A_{Bp}$ and $A_{Rp}$ using \gaia\ EDR3 extinction laws\footnote{https://www.cosmos.esa.int/web/gaia/edr3-extinction-law}.

We compare this photometry with the low-mass stellar evolution models of \citet{Baraffe2015}, the SPOTS models of \citet{Somers2020}, and standard and magnetic versions of the Dartmouth models \citep{Feiden2014,Feiden2016}.  It is now widely appreciated that magnetic fields of low-mass pre-main sequence stars can inflate the stars, making them appear younger than they actually are \citep{Feiden2016}.  The SPOTS models attempt to account for both the magnetic and two-temperature effects of starspots with different values of spot fractional coverage.  The Dartmouth models especially computed for this study assume the solar composition described by \citet{Grevesse2007}. Physics included in these models are described in \citet{Feiden2014} and references therein, and are consistent with Dartmouth models used in previous studies of young stars \citep[e.g.,][]{Malo2014,Stassun2014,Feiden2016}. Magnetic model physics are described in \citet{Feiden2012} with subsequent modifications outlined by \citet{Feiden2016}. In addition, the Dartmouth magnetic models evolve with a surface magnetic field strength in equipartition with the photospheric gas pressure during each model time step. Photometric magnitudes were calculated using synthetic spectra from MARCS model atmospheres \citep{Gustafsson2008} with \gaia\ DR2 zero-points established by \citet{Casagrande2018}. The same synthetic transformations were used for standard and magnetic models.

\begin{figure}
	\includegraphics[width=\columnwidth]{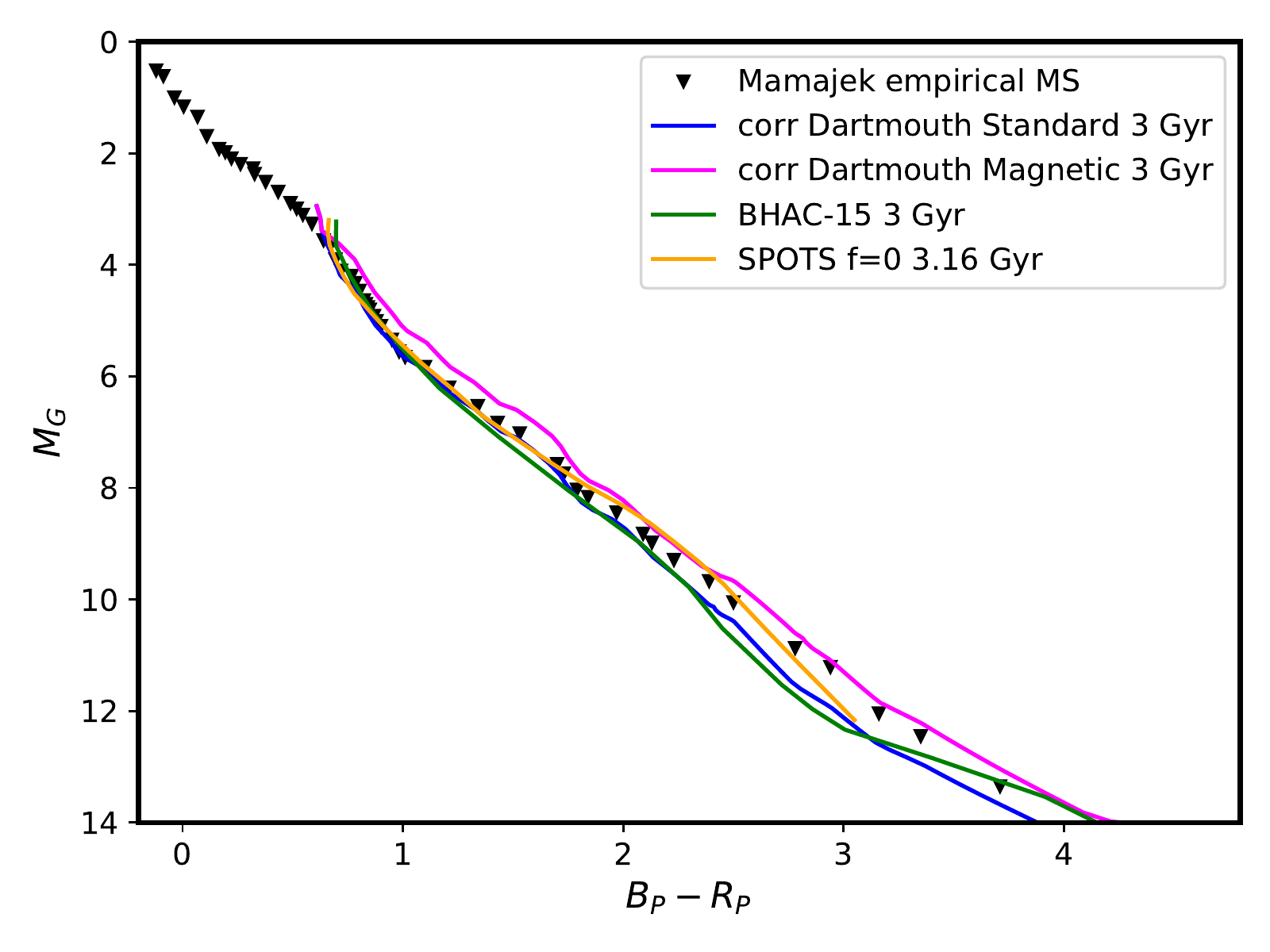}
    \caption{Empirical \gaia\ color-magnitude diagram of main-sequence stars as described by \citet{Pecaut2013} (black diamonds) compared to 3 Gyr-isochrones from different models used to estimate the age of the Carina moving group.}
    \label{fig:cmd_mamajek}
\end{figure}

The high-precision photometry of \gaia\ has revealed discrepancies between model-predicted magnitudes and colors for the coolest stars.  To avoid the inaccuracies of synthetic values based on model spectra, \citet{Somers2020} related \teff\ to colors and bolometric corrections using the empirical values compiled by \citet{Pecaut2013}.  We follow the same procedure with the BHAC-15 and Dartmouth models.  Figure \ref{fig:cmd_mamajek} compares 3 Gyr model isochrones adjusted in this way with the empirical values of \citet{Pecaut2013}, demonstrating generally good agreement.

\begin{figure}
	\includegraphics[width=\columnwidth]{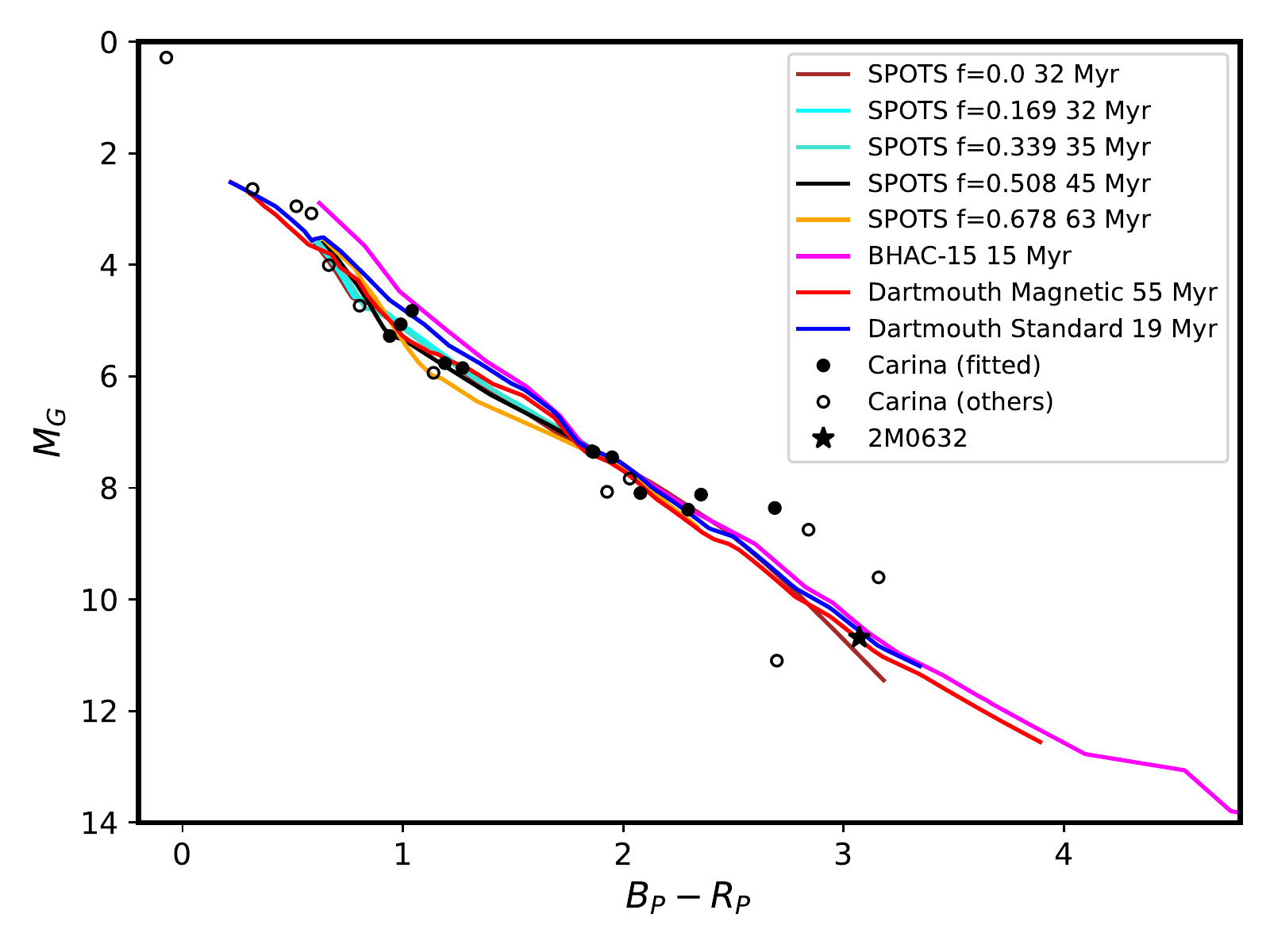}
    \caption{\gaia\ color-magnitude diagram of Carina members and best-fit isochrones from different models.  Photometry has been corrected for interstellar extinction/reddening.  Black filled points were used in the fits; open points were excluded from the isochrone fitting based on RUWE (binarity), anomalous extinction, or outside the color range covered by the models.}
    \label{fig:cmd}
\end{figure}

Following \citet{Naylor2006} and \citet{Naylor2009} we fit isochrones by minimizing the $\tau^2$ metric, a 2-d version of the $\chi^2$ goodness-of-fit parameter that is a summation over $N$ stars:    
\begin{equation}
    \tau^2 = -2 \sum_{i=1}^{N}\log \int dC \int dm \, \rho(C,m)\exp \left[-\frac{\left(C-C_i\right)^2}{2\sigma_{C}^2}-\frac{\left(m-m_i\right)^2}{2\sigma_{m}^2}\right],
    \label{eqn:tau-squared}
\end{equation}
where $\rho(c,m)$ is the model-predicted distribution of stars in the color-magnitude ($C$-$m$) plane.  Since brightness and color are single-valued functions of mass for a given (solar) metallicity, we collapsed the 2-d integral over $C$ and $m$ into a single integral with stellar mass $M_\star$ along the isochrone.  For $\rho(M_\star)$, we adopted an initial mass function of $M_\star^{-\alpha}$ with $\alpha=2.3$ \citep{Chabrier2000}.  Isochrone-based masses for each star are calculated as the expectation value:
\begin{equation}
\langle M_\star \rangle  = \frac{\int dM_\star M_\star \rho(M_\star)\exp \left[-\frac{\left(C(M_\star)-C_\star\right)^2}{2\sigma_{C_\star}^2}-\frac{\left(m(M_\star)-M_\star\right)^2}{2\sigma_{M_\star}^2}\right]}{\int dM_\star \rho(M_\star)\exp \left[-\frac{\left(C(M_\star)-C_\star\right)^2}{2\sigma_{C_\star}^2}-\frac{\left(m(M_\star)-M_\star\right)^2}{2\sigma_{M_\star}^2}\right]}
    \label{eqn:mass}
\end{equation}
Figure \ref{fig:cmd} plots the best-fit isochrones from the different model sets.  A range of ages are found (15-63 Myr) depending on whether magnetic field effects and spots are included.  The two best-fitting models are the SPOTS model for $f=0$ and 32 Myr age, and the Dartmouth magnetic model with 55 Myr age. 

There are three systems which are significantly ($>$1 mag) more luminous than predicted by the best-fit model isochrones.  AL\,442 (2MASS J06112997-7213388) does not have a published RUWE value but has been resolved into a 0\arcsec.16-separation binary by speckle imaging \citep{Janson2012}. 2MASS J09315840-6209258 has RUWE=1.3, only marginally suggestive of binarity; two measurements of RV agree \citep{Malo2013,Schneider2019}.  2MASS J09180165-5452332 (RUWE = 5.32) was also resolved in speckle imaging \citep{Janson2012}.  However, binarity alone cannot explain the $>1$~mag offset of these stars above the best-fit isochrones (Fig. \ref{fig:cmd}).  These could conceivably be higher-order systems.  These could also be younger ($<$15 Myr) interlopers but the high probability of membership disfavors this explanation.

\begin{figure}
    \centering
    \includegraphics[width=\columnwidth]{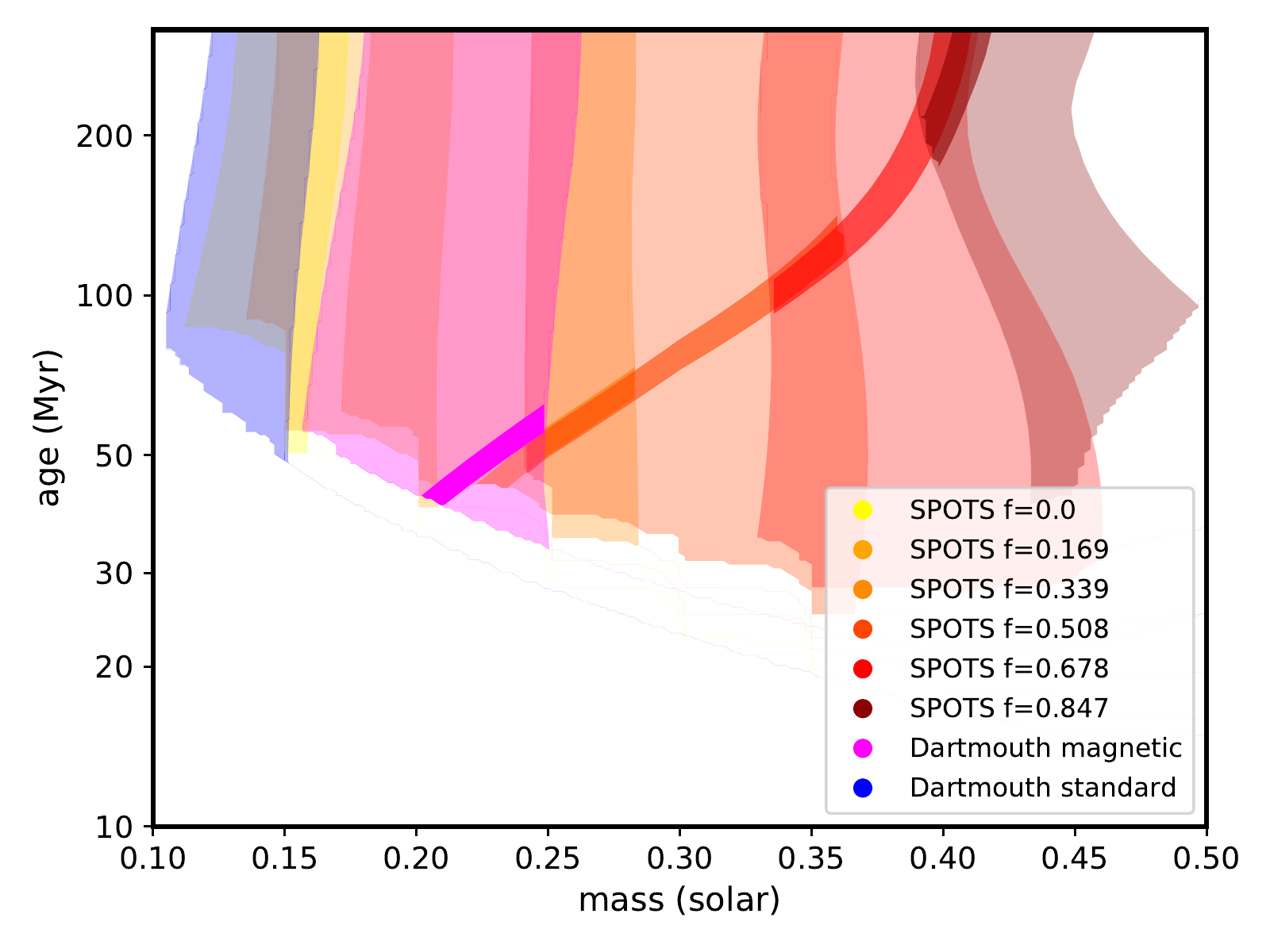}
    \caption{Li abundance constraints on the age of \thestar\ (alone) using different stellar evolution models.  The lines demarcate the boundary above which Li is depleted below the observed upper limit; the lightly shaded areas the zones of those regions in which model \teff\ is within 75K of the inferred value (3100K) and the more intensely shaded areas within are those where model $K_s$-band luminosity is within 2$\sigma$ of calculated value.}
    \label{fig:lithium}
\end{figure}

Because Li is rapidly depleted in cool star photospheres, it is a useful constraint on the age of young moving group members.  The Li doublet at 6708\AA\ was not detected in our optical spectrum of \thestar\ (Fig. \ref{fig:soar}) and this sets a minimum model age for the star.  A comparison with simulated lines using the curve of growth for \teff=3100K and $\log g = 4$ of \citet{Palla2007} shows that the $\log$ abundance $A(Li)$ (relative to H=12) is $<0$, or  $<6.3\times 10^{-4}$ of the interstellar value.  We find that SPOTS models with spot fraction coverage of 0 or 0.17 cannot simultaneous satisfy the constraints on \teff\ (3100$\pm$75K), absolute $K_s$-band magnitude ($6.642 \pm 0.023$) and $A(Li) <0$ (Fig. \ref{fig:lithium}).   Higher spot fractions can satisfy all three constraints, but only for model ages of at least 43 Myr.  Likewise, neither the standard Dartmouth models or the models of \citet{Baraffe2015} can simultaneous reproduce the luminosity, \teff\ and Li depletion, but the magnetic model can, presuming an age of 40-60 Myr (Fig. \ref{fig:lithium}).  

\section{Summary and Discussion}
\label{sec:discussion}

\subsection{What causes the dimming of \thestar?}

An unparalleled record of two years of precision \tess\ photometry of \thestar\ containing $>300$ dimming events provides a long-term, statistical insight into the mechanism or mechanisms responsible for occultation of the star.  Multiple scenarios have been proposed to explain ``dipper" stars, some of which appear to be ruled out by these observations.  First, the dimming behavior is persistent and the pattern is largely unchanged on time-scales of years, inconsistent with explanations involving the disruption of and dust production by a small number of disintegrating planetesimals or ``exocomets".  There is a possible increase in the occurrence of events (Fig. \ref{fig:triangle}), reflected in an increase in the \citet{Cody2014} asymmetry parameter $M$ from 2.4 to 11.6 from Year 1 to Year 2, although $M$ is heavily biased by the few deepest events.  

Second, all dips have $\tau_{50} > 1$~hr and the great majority are $>$2 hr, and if the occulting material is similar in scale to the star, this sets an upper limit on the orbital speed and lower limit on the orbital period of the material.  Assuming that the actual duration of the occultation is twice $\tau_{50}$, the mass and radius of the star then imply an orbital period of $P \sim 0.75 \rho_\star \tau_{50}^3$, where $P$ is in days, $\rho_\star$ is the stellar density in solar units, and $\tau_{50}$ is in hours.  Since  $\rho_\star \approx 2.8\rho_{\odot}$ (Table \ref{tab:params}), the implied orbital period is $\gtrsim$1.5 days.  This would suggest that the inner edge of the disk coincides with the co-rotation radius ($R_{\rm cor}$, P=1.86 days, Sec. \ref{sec:tess_analysis}).  The equilibrium black-body temperature at $R_{\rm cor}$ is $\sim$700K.  This is consistent with the excess emission detected by \wise\ at 3-5$\mu$m (Fig. \ref{fig:sed}) and thus this excess could be from the inner disk.  But we caution that the three relevant points in the SED (at 2.2, 3.4, and 4.6 \micron) only poorly constrain the temperature, and inner disk emission is probably not described by a single-temperature black body.  If ISM-like reddening obtained from our LCOGT photometry precludes occultation by dust at these longer IR wavelengths as an explanation, the correlated variability at 3.4 and 4.6 $\mu$m suggests that this region of the disk could have variable emission.    

The stochasticity of the dimming signal ($Q = 0.90$) and the range of dip durations (Fig. \ref{fig:triangle}) suggests that the occulting structures are not organized and long-lived, but are short-lived and occur over a region of the inner disk, in contrast with the more persistent quasi-periodic behavior of AA Tau \citep{Bouvier2003}, However, the significant clustering of dips with rotational phase and presumably the distribution of spotted regions on the star (Fig. \ref{fig:phased}) suggsts a connection with the stellar magnetic field analogous to but weaker than AA Tau-like magnetically-funneled accretion \citep{Bouvier2003}.  The non-uniform distribution of rotational phase of the dips, with more dips occurring in the minimum of the light curve when the more spotted side of the star is visible (Fig. \ref{fig:phased}), could be a manifestation of dusty accretion controlled by an asymmetric stellar magnetic field.   Asymmetries (skewness) in the light curve could be produced by the rotating magnetic field of the star acting on the gas (and dust) once it decouples from the disk.  Structures outside (inside) the co-rotation radius will experience positive (negative) torques that accelerate gas forward (backwards) and produce leading (trailing) tails with positively (negative) skewed light curves.  In this scenario, the distribution of skewness values (Fig. \ref{fig:triangle}) would be the result of structures appearing over a range of radii encompassing the co-rotation radius, with a majority interior to that distance.

Following \citet{Bodman2017}, we related the accretion rate $\dot{M}$ in the disk to extinction $A_T$ in the \tess\ band produced by an accretion stream extending from the inner edge of the disk, assuming a vertical cylindrical geometry, and extinction coefficient $\mu_T$ (magnitudes per unit gas mass surface density) and a disk inner edge at the co-rotation radius $R_{\rm co}$:
\begin{equation}
\label{eqn:extinction1}
A_T = \frac{\mu_T \dot{M}}{2m_p R_\star v_s}\left[\frac{\tau}{ P_{\rm cor}^{1/3}\tau_{\rm ff}^{2/3}} - 1 \right]^{-1},
\end{equation}
where $m_p$ is the proton mass, $v_s$ is the velocity along the stream, $P_{\rm cor}$ is the Keplerian period at the inner edge of the disk, which we assume to be at the co-rotation radius \citep{Stauffer2017}, and the free-fall time is defined as $\tau_{\rm ff} = \left(\pi^2G\rho_\star/3\right)^{-1}$.  The final factor relates the duration of the dip $\tau$ to the width of the accretion stream.  We related $v_s$ to the magnetic field pressure at the disk inner edge by assuming that at this location the vertical flow is driven by magnetic pressure such that $v_s \sim \sqrt{B^2/(2\rho)}$, where $\rho$ is the gas density \citep[e.g.,][]{Bessolaz2008}.  Assuming a pure dipole field that extends from the surface with strength $B_\star$, the field at the disk inner edge is $B \approx B\star (R_\star/R_{\rm cor})^{-3}$.   We used a value $\mu_T = 3.0 \times 10^{-22}$~cm$^{-2}$ for the ISM found by combining $N_H/A_V = 2.21 \pm 0.09 \times 10^{21}$ cm$^{-2}$ from \citet{Guver2009}, $A_T = 2.06 E(B-V)$ from \citet{Stassun2018} and the standard ISM extinction law of $A_V = 3.1E(B-V)$:  
\begin{multline}
    A_T \approx 9.3 \frac{\mu_s}{\mu_{\rm ISM}}  \left(\frac{\dot{M}}{10^{-10}M_{\odot}~yr^{-1}} \right)^{2} 
     \left(\frac{R_\star}{R_{\odot}}\right)^{-3} \left(\frac{B_\star}{{\rm 1~ kG}}\right)^{-2} \left(\frac{P_{\rm rot}}{\rm 1~ day}\right)^4 \\ \times \left(\frac{\rho_\star}{\rho_{\odot}}\right)^2 \left[\frac{\tau}{6.7~{\rm hr}}\left(\frac{\rho_\star}{\rho_{\odot}} \right)^{1/3}-1\right]^{-1}
     \label{eqn:extinction2}
\end{multline}
Adopting the parameters from Table \ref{tab:params}, the $H_\alpha$/He~I EW-based accretion rate, $\tau$=8 hr, and $B_\star = 3$~kG \citep[typical for young, rapidly-rotating M dwarfs,][]{Reiners2022}, yields extinction $A_T \sim 0.2$, remarkably consistent with the distribution of dip depths (Fig. \ref{fig:triangle}).  We note, however, that our estimate of $A_T$ is sensitive to the chosen values of the stellar properties $B_\star$, $M_\star$, and $R_\star$.  

Finally, the mechanism producing the occultations of \thestar\ are operating in a disk that is significantly older than the those of most known ``dipper" stars, e.g. in the 1-5 Myr-old Taurus, 3 Myr-old $\rho$ Ophiucus, 3-5 Myr-old Lupus, and $\sim$10 Myr-old Upper Scorpius star-forming regions \citep{Rodriguez2017,Roggero2021,Bredall2020,Ansdell2016a}.  Noteworthy in this respect is the fact that WISE\,J080822.18-644357.3, another Carina member that hosts a long-lived disk \citep{Murphy2018}, is also a ``dipper" \citep{Silverberg2020}.  Either the mechanism is invariant with disk evolution, which seems inconsistent with the relative paucity of decrease in the dipper phenomenon among more evolved disks \citep{Ansdell2016a}, or disk evolution was slowed or halted around \thestar.   Indeed, if the dust responsible for the variable dimming and reddening of \thestar\ is indeed ISM-like, a reservoir where this material can escape significant processing for tens of Myr is required.

\subsection{The age of \thestar\ and the Carina group}

\thestar\ is undoubtedly a very young M dwarf star; its position and space motion support membership in the Carina young moving group and should thus be approximately the same age as the group or progenitor cluster.  Our comparison with model stellar isochrones and the absence of Li suggest an age of at least 40 Myr and perhaps as much as 60 Myr for the Carina group and \thestar, supporting earlier estimates \citep[e.g.,][]{Bell2015} but not supporting more recent proposals for an age of $\sim$15 Myr \citep{Booth2021}.  

Our work highlights three reasons why assigning  a definitive age to the group is challenging.  First, the group is dispersed and sparse, and relatively few faint (M dwarf) members have been confirmed with RV measurements.  These M dwarfs evolve more slowly on the pre-main sequence than solar-mass stars, and are thus most informative about the age of the group.  By the age of Carina, solar-mass stars are at or near the zero-age main sequence.  Second, models can include the effects of magnetic fields and starspots and the degree of these effects (e.g., the spot coverage fraction) is significant but poorly constrained, and likely varies between stars.  Third, the precision of \gaia\ photometry has revealed significant discrepancies with model predictions of colors and/or luminosities that are incompletely addressed by the empirical ``bandaging" performed by \citet{Somers2020} and this work.

\subsection{The nature of long-lived disks}

\begin{figure}
	\includegraphics[width=\columnwidth]{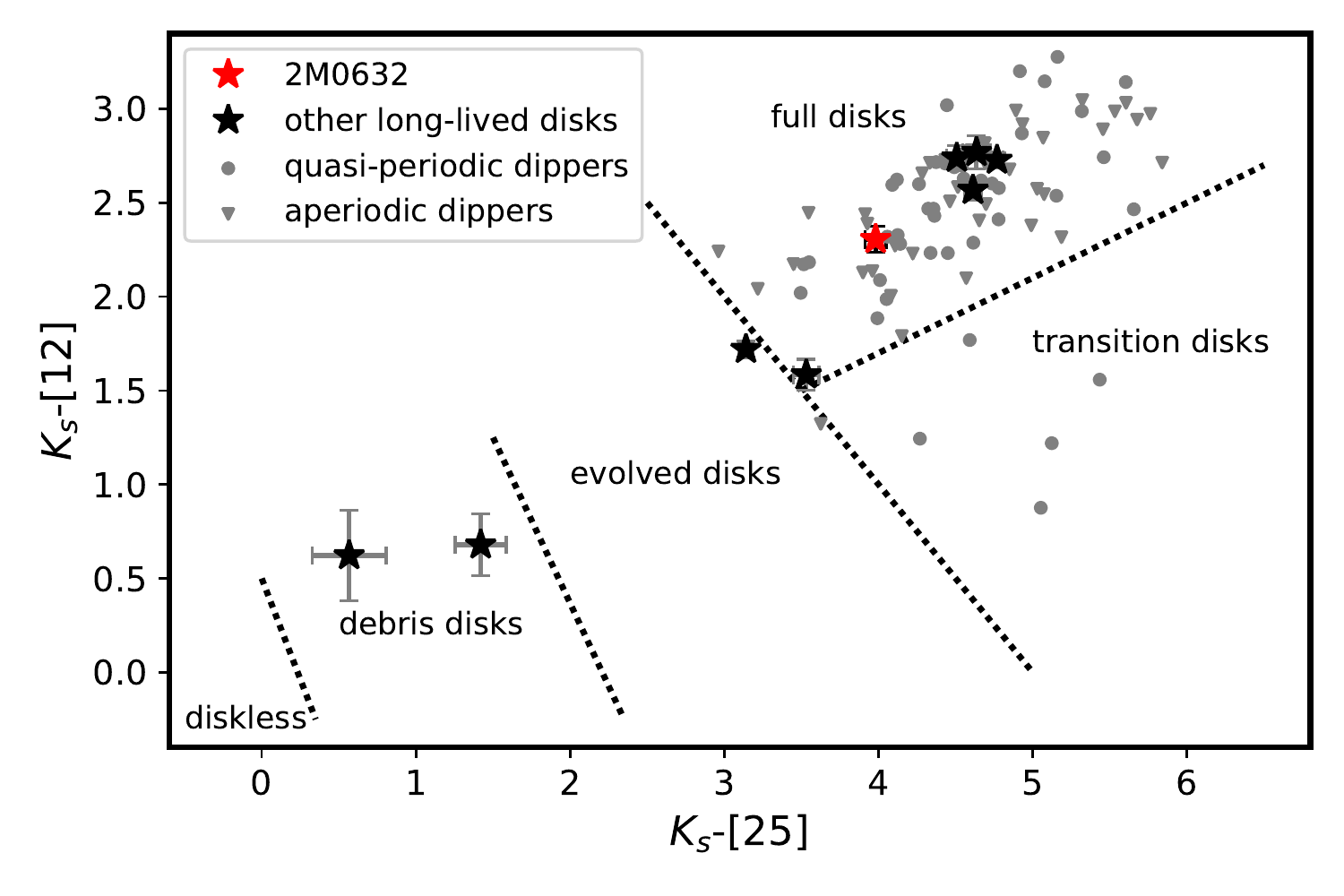}
    \caption{Infrared color-color diagram of emission from disks of dipper stars based on \wise\ 12- and 25-\micron\ and 2MASS $K_s$ (2.2 \micron) photometry.  Carina member \thestar\ is marked by a red star and black stars are other long-lived disks cataloged in \citet{Silverberg2020}.  Grey points are dipper members of the 3-10\,Myr-old $\rho$ Ophiuchus and Upper Scorpius star-forming regions identified by \citet{Cody2018} and distinguished by their light curve morphology.  Dashed lines are the approximate boundaries between the different disk types as described by \citet{Luhman2012}.}
    \label{fig:excess}
\end{figure}

If a $\gtrsim$40 Myr age for Carina does hold up, then \thestar\ joins several other systems (including others from Carina) with long-lived disks.  We plot \thestar\ and previously recognized examples \citep{Silverberg2020} in the WISE infrared color-color diagram (Fig. \ref{fig:excess}). 2MASS\,J05082729-2101444 and SCR\,J0103-5515, have excess infrared emission more consistent with debris disks than with primordial T Tauri-like disks  and in that sense are not exceptional.  2MASS\,J05082729-2101444 is a confirmed M dwarf member of the $\sim$25 Myr-old $\beta$ Pictoris Moving Group (BPMG) and has H$\alpha$ emission only marginally consistent with accretion \citep{Schneider2019,Lee2020}.  SCR\,J0103-5515 is a member of the $\sim$45 Myr-old Tucana-Horologium cluster \citep{Malo2013} and a potential debris disk candidate \citep{Binks2017}.  Two other systems, 2MASS\,J04463413-2627559 and 2MASS\,J09490073-7138034, are borderline full/evolved/transition disks \citep{Silverberg2020}.  

The four other long-lived disks that cluster in the ``full" disk part of the infrared color-color space (fellow Carina member WISE\,J080822.18-644357.3, $\sim$60 Myr-old Argus member 2MASS J15460752-6258042 \citep{Lee2020},  and BPMG binary LDS\,5606A+B \citep{Rodriguez2014}  all appear to be single stars or in the last case a wide ($\rho$=2200\,au) binary.  This is consistent with considerable data that T Tauri-like disks around the individual components of binary systems (S-type disks) tend to be shorter-lived \citep[e.g.,][]{Kraus2012a}.  P-type disks that surround a close ($\lesssim$1 AU) binary could be long-lived \citep{Alexander2012,Ansdell2020}, as well as disks shepherded between the inner and outer components of a hierarchical systems \citep{Ronco2021}, but these cases do not seem to apply here.  Curiously, all four long-lived debris or transitional disks are are in close ($\lesssim$1" binaries or candidate binaries based on RUWE (in the case of 2MASS~J05082729-2101444), suggesting that the presence of a binary companion does not preclude the persistence of a partially cleared disk or substantial debris disk.

While it is possible these long-lived disks are experiencing arrested development \citep{Silverberg2020}, there is suggestive evidence of evolution away from the gas-rich state characteristic of protoplanetary disks.  ALMA observations detect cold dust but no (CO) gas in the disk around WISE 0808 \citep{Flaherty2018}.  This could indicate that a more debris-disk-like state, or perhaps reflect the short lifetime of CO against photodissociation over the prolonged lifetime of such disks.

\subsection{What does the existence of long-lived disks imply?}
\label{sec: implications}

Given that the phenomenon of long-lived ($\gg$10\,Myr) disks around very low-mass stars is increasingly secure, we must ask whether these represent the tail of an underlying distribution, a manifestation of a fundamental difference between disks around solar-type stars and poorly-studied M dwarfs, or anomalies  And what are the implications for our understanding of disk evolution, planet formation, and the potential differences between M dwarfs and their planets, and solar-type stars?  

Simple models of disk evolution offer insight into the parameter space in which disks may persist for 10s of Myr.  \citet{Coleman2020} found that a combination of elevated disk mass, low radial transport (i.e, low $\alpha$ parameter) \emph{and} exceptionally low photoevaporation rates were all required for disks to persist over the observed ages.  They argued that disks can only persist in low-UV environments embedded in cloud regions at the outskirts of star-forming regions.  In contrast, \citet{Wilhelm2022} find that internal (central-driven) photoevaporation is crucial and that, because lower-mass stars have lower X-ray/EUV luminosities \citep[which scale with bolometric luminosity, e.g.,][]{Wright2018,France2018}, $<0.6$\msun\ stars can retain their disks for $\sim$50 Myr.  The crucial parameter responsible for these divergent model outcomes is the efficiency of photoevaporation by the external FUV-field, a process which for M dwarfs could dominate over loss driven by the central star, but which could be diminished for stars formed in the low-FUV environments of clusters that are gas-rich and/or lack OB stars.  These conclusions also hinge on the reliability of models of such mass loss \citep{Haworth2018}.   While it is tempting to relate the existence of multiple long-lived disks in the Carina moving group to its low membership, \citet{Gagne2021} has proposed that the group is a tidally stripped fragment of the Platais 8 cluster and thus could have experienced a denser stellar environment and higher FUV irradiance in the past.

The consequences of a long-lived disk for planet formation have not been thoroughly explored, but the canonical theory offers some hypotheses.   Because the timescale of the assembly of rocky cores by giant impacts scales with Keplerian orbital period, this formation channel should be little affected close ($\ll$1 au) to the star \citep{Zawadzki2021} but could be enhanced at greater separation.  Provided sufficiently massive cores form, long-lived gaseous disks should permit accretion of more massive H/He envelopes \citep{Ribas2015}.  Other affects arise from migration of solids through a gas disks: The slightly non-Keplerian rotation of a pressure-supported gas disk imparts aerodynamic drag on cm- to meter-size bodies, causing them to migrate and concentrate inwards and at any pressure maxima.    Torques from a gas disk also cause ``Type I" migration of planets, and the uninterrupted migration of multiple planets is predicted to result in compact systems of resonant ``chains", with planets on highly circular, co-planar orbits starting at the disk's inner edge \citep{Izidoro2017}.  Thus a long-lived disk might be expected to produce very rich, compact systems of planets.  Although the planets may have originally captured H/He from the disk into extended envelopes, their proximity to the young, active star is expected to rapidly evaporate those atmospheres \citep{Owen2019}.  These attributes recall the exceptional system TRAPPIST-1, where 7 Earth-size planets orbit close to an ultra-low-mass M dwarf \citep{Lienhard2020}, and it is conceivable that \thestar\ is a forerunner of similar systems.

\section*{Acknowledgements}

We thank Sean Points of the NSF NOIRLab for carrying out the TripleSpec-4.1 observations and data reduction.  E.G. and S.N. acknowledge support by NASA grants 80NSSC19K0587 (Astrophysics Data Analysis Program) and 80NSSC19K1705 (\tess\ Guest Observer Cycle 2).   A.W.M. was supported through NASA’s Exoplanet Research Program (XRP; 80NSSC21K0393).  B.R-A acknowledges funding support from FONDECYT Iniciaci\'{o}n grant 11181295 and ANID Basal project FB210003.  This paper includes data collected by the \tess\ mission. Funding for the \tess\ mission is provided by the NASA's Science Mission Directorate.  This work has made use of data from the European Space Agency (ESA) mission {\it Gaia} (\url{https://www.cosmos.esa.int/gaia}), processed by the {\it Gaia} Data Processing and Analysis Consortium (DPAC, \url{https://www.cosmos.esa.int/web/gaia/dpac/consortium}). Funding for the DPAC has been provided by national institutions, in particular the institutions participating in the {\it Gaia} Multilateral Agreement.This work makes use of observations from the Las Cumbres Observatory global telescope network.  Based in part on observations obtained at the Southern Astrophysical Research (SOAR) telescope, which is a joint project of the Minist\'{e}rio da Ci\^{e}ncia, Tecnologia e Inova\c{c}\~{o}es (MCTI/LNA) do Brasil, the US National Science Foundation’s NOIRLab, the University of North Carolina at Chapel Hill (UNC), and Michigan State University (MSU). This research has made use of the archive of the European Southern Observatory, and the NASA/IPAC Infrared Science Archive, which is funded by the National Aeronautics and Space Administration and operated by the California Institute of Technology. This research or product makes use of public auxiliary data provided by ESA/Gaia/DPAC/CU5 and prepared by Carine Babusiaux.  This publication makes use of VOSA, developed under the Spanish Virtual Observatory project supported by the Spanish MINECO through grant AyA2017-84089. VOSA has been partially updated by using funding from the European Union's Horizon 2020 Research and Innovation Programme, under Grant Agreement no. 776403 (EXOPLANETS-A) .   We used NASA's Astrophysics Data System Bibliographic Services, the Centre de Donn\'{e}es astronomiques de Strasbourg, {\tt Astropy} \citep{Astropy2013}, and {\tt Scipy} \citep{Scipy2019}.\\

\textbf{Data Availability Statement:} \tess\ data is publicly available through MAST archive at STScI.  \gaia\ data is publicly available through the Centre de Donn\'{e}es astronomiques de Strasbourg (CDS).  LCOGT data are either publicly available from its archive or from the authors upon request.  All ESO data are publicly available from its archive.  All other photometry is available through the Virtual Observatory or the individual catalogs maintained at the CDS.  The SOAR Goodman Echelle and TripleSpec-4.1 spectra are available upon request from the authors.





\bsp	
\label{lastpage}
\end{document}

%% file: newcommands.tex
\newcommand{\rsos}{Royal Soc. Open Sci.}

\newcommand{\halpha}{H$\alpha$}

\newcommand{\galex}{\emph{GALEX}}
\newcommand{\tess}{\emph{TESS}}

\newcommand{\gaia}{\emph{Gaia}}
\newcommand{\ktwo}{\emph{K2}}
\newcommand{\jwst}{\emph{JWST}}
\newcommand{\kepler}{\emph{Kepler}}

\newcommand{\wise}{\emph{WISE}}

\newcommand{\kms}{\mbox{km\,s$^{-1}~$}}

\newcommand{\msun}{M$_{\odot}$}

\newcommand{\lsun}{L$_{\odot}$}

\newcommand{\teff}{\ensuremath{T_{\rm eff}}}

